\documentstyle[11pt,aaspp,psfig]{article}
\def\gs{\mathrel{\raise0.35ex\hbox{$\scriptstyle >$}\kern-0.6em 
\lower0.40ex\hbox{{$\scriptstyle \sim$}}}}
\def\ls{\mathrel{\raise0.35ex\hbox{$\scriptstyle <$}\kern-0.6em 
\lower0.40ex\hbox{{$\scriptstyle \sim$}}}}

\def\arcsper{\ifmmode \rlap.{''}\else $\rlap{.}''$\fi}
\def\arcmper{\ifmmode \rlap.{'}\else $\rlap{.}'$\fi}

\def\MD{\hbox{$T-\Sigma$~}}
\received{December 26, 1996}
\revised{July 6, 1997}
\accepted{July 15, 1997}

\begin{document}

\title{EVOLUTION SINCE Z = 0.5 OF THE MORPHOLOGY-DENSITY \\
RELATION FOR CLUSTERS OF GALAXIES
\footnote{Based on observations 
obtained with the NASA/ESA Hubble Space Telescope
which is operated by STSCI for the Association of Universities
for Research in Astronomy, Inc., under NASA contract NAS5-26555.}}
\vspace{4mm}
\author{
Alan Dressler\altaffilmark{1},
Augustus Oemler, Jr.\altaffilmark{1},
Warrick J.\ Couch\altaffilmark{2},
Ian Smail\altaffilmark{3}\footnote{Visiting Research Associate at the Carnegie
Observatories.},
Richard S.\ Ellis\altaffilmark{4},
Amy\ Barger\altaffilmark{4},
Harvey Butcher\altaffilmark{5},
Bianca M. Poggianti\altaffilmark{4,5,6} \&
Ray M.\ Sharples\altaffilmark{3} 
}

\affil{\scriptsize 1) The Observatories of the Carnegie Institution, 813 Santa Barbara St., Pasadena, CA 91101-1292}
\affil{\scriptsize 2) School of Physics, University of New South Wales, Sydney 2052, Australia}
\affil{\scriptsize 3) Department of Physics, University of Durham, South Rd, Durham DH1 3LE, UK}
\affil{\scriptsize 4) Institute of Astronomy, Madingley Rd, Cambridge CB3 OHA, UK}
\affil{\scriptsize 5) Kapteyn Instituut, PO Box 800, 9700 AV Groningen, The Netherlands}
\affil{\scriptsize 6) Royal Greenwich Observatory, Madingley Road, Cambridge CB3 0EZ, UK}

\begin{abstract}
Using traditional morphological classifications of galaxies in 10
intermediate-redshift (z $\sim$ 0.5) clusters observed with WFPC-2 on
the Hubble Space Telescope, we derive relations between morphology and
local galaxy density similar to that found by Dressler for low-redshift
clusters.  Taken collectively, the "morphology-density" relationship,
\MD, for these more distant, presumably younger clusters is
qualitatively similar to that found for the local sample, but a
detailed comparison shows two substantial differences: (1) For the
clusters in our sample, the \MD relation is strong in centrally
concentrated ``regular'' clusters, those with a strong correlation of
radius and surface density, but nearly absent for clusters that are
less concentrated and irregular, in contrast to the situation for low
redshift clusters where a strong relation has been found for both. (2)
In every cluster the fraction of elliptical galaxies is as large or
larger than in low-redshift clusters, but the S0 fraction is 2-3 times
smaller, with a proportional increase of the spiral fraction.

Straightforward, though probably not unique, interpretations of these
observations are (1) morphological segregation proceeds hierarchically,
affecting richer, denser groups of galaxies earlier, and (2) the
formation of elliptical galaxies predates the formation of rich clusters,
and occurs instead in the loose-group phase or even earlier, but S0's are
generated in large numbers only after cluster virialization.

\end{abstract}


\sluginfo
\newpage

\section{Introduction}

The observation that galaxy populations in clusters differ from those
in the field goes back at least as far as Hubble \& Humason (1931).
Both Spitzer \& Baade (1951) and Gunn \& Gott (1972) suggested that
dynamical processes within clusters might be responsible for
transforming populations of spiral galaxies into S0's.  In a study of
10 rich clusters, Oemler (1974) found that the cluster population was a
function of both cluster structure and location within the cluster.
Regular, centrally concentrated clusters, whose appearance suggests
that they are dynamically relaxed, have large populations of E's and
S0's, small numbers of spirals, and strong radial population gradients,
with the ratio of early-to-late-type galaxies increasing towards the center.
Irregular, unrelaxed-looking clusters have populations that are much more spiral-rich and show no signs of radial gradients. Oemler suggested,
following Gunn and Gott, that the high S0/spiral ratio in relaxed
clusters was due to dynamical processes within the clusters, but that
the enhanced abundance of E's in the most centrally concentrated
clusters, which are often dominated by cD galaxies, was due to an
enhanced formation rate of ellipticals at early times in the densest
environments that would later become the cores of rich clusters.

Dressler (1980a, hereafter D80) used data for $\sim$6000 galaxies 
(Dressler 1980b, hereafter DCAT80) to investigate correlations between
morphological type, cluster properties, and spatial distribution in 55
low-redshift rich clusters.  Dressler correlated the spiral, S0, and
elliptical fractions with the {\it local} surface density of galaxies,
defined in a rectangular area containing the 10 nearest neighbors. He
found a smooth, monotonic relation, now commonly known as the {\em
morphology--density relation}, actually a correlation with surface
density, which we will refer to here as the \MD relation.  Dressler
concluded that, to first order at least, the \MD relationship was
universal, that is, representative of every cluster in the sample,
regardless of its global properties.  (Indeed, other workers (e.g.
Geller \& Beers 1982) have claimed that it extends beyond cluster
environments into the general field population.) Of particular
importance was Dressler's finding that, while irregular clusters showed
no radial segregation of cluster populations, as Oemler had shown, they
did display a \MD relation as strong as that of regular, concentrated
clusters.  Of course, the finite number of galaxies in each cluster
prevents the testing of this hypothesis on a case-by-case basis, so
that this conclusion was reached by combining together the data for
regular, centrally-concentrated clusters and comparing them to those of
irregular, clusters of low central concentration.

Both Oemler's and Dressler's portrayals of cluster populations are
consistent with environment playing a critical role in determining
present galaxy populations. Indeed, for relaxed, centrally-concentrated
clusters the two descriptions are equivalent, because there is a
one-to-one correspondence between radial position and local density in
such clusters. However, the difference for irregular clusters is
significant. Were irregular clusters to show little segregation by type, 
it would be reasonable to conclude that evolutionary processes which 
transform galaxies only work in the environment of a massive, relaxed 
system. On the other hand, finding local correlations between 
environment and populations even in small, cold subsystems suggests that
much more local processes are at work.

The purpose of this paper is to investigate these issues in clusters at
higher redshift, $z \sim 0.5$, in order to trace the evolution of
galaxy populations in clusters, specifically the processes that might
alter galaxy types and/or be responsible for their spatial distribution
within clusters.  This paper is one of a series analyzing the data
obtained by the ``MORPHS'' group using the WFPC-2 on the Hubble Space
Telescope.  We have imaged 11 fields in 10 clusters of galaxies in the
redshift range $0.36 < z < 0.57$ and cataloged the positions,
photometric data, and morphological classifications for 1857 objects
brighter than $R = 23.5$ or $I = 23.0$, as described in Smail et
al.\ (1996a, hereafter, S97a).  Although the addition of further
parameters derived from spectroscopy of these galaxies, such as cluster
or field membership, internal dynamics, stellar population, and cluster
dynamics, are likely to be revealing as to the evolutionary state of
these populations, photometric/morphological information alone allows a
simple and important comparison with the properties of present day
clusters.  This comparison offers clues as to how clusters of galaxies
came to hold their atypical complements of galaxy types.

The paper is organized as follows:  In section 2 we briefly review the
data used for this study.  In Section 3 we revisit the relation between
morphology and surface density introduced in D80. We include a short
discussion of subsequent challenges to the original work, which is
further explored in the Appendix.  Section 4 gives our results for the higher-redshift sample, and in Section 5 we suggest some possible 
conclusions about galaxy evolution in clusters that might be drawn from 
our results.

\section{A Brief Review of the Data}

As described in S97a, our data consist of images of 11 fields of 10 rich
clusters, obtained with the WFPC-2 on the Hubble Space Telescope.  The
exposures are typically 4 to 6 orbits each of $\sim$2000 s duration.
In a few cases fields were observed in more than one band, but for the
purposes of this paper only F814W or F702W exposures are involved,
since these alone were used for morphological classification.  By
inspecting the images in what is essentially the rest-frame B-band, we
have tied our system to the Revised Hubble and de Vaucouleurs schemes,
which are traditionally based on images in the B-band --- this mitigates legitimate concerns about type-dependent {\it k} corrections.  The
morphological classifications used here follow the traditional method
of visual inspection of images. As discussed in S97a, at least two,
and often three or four of us assigned a morphological type for an 
image, with a high degree of repeatability and consistency.

The clusters in our sample were not selected through any systematic
criteria, but represent many of the best examples of well-studied
clusters at intermediate redshift.  As such, they are all rich
(populous) clusters, though with a range in richness of a factor of
three.  They also represent a range in dynamical state, as evidenced by
the regularity of the spatial distributions of their galaxies: the
clusters range in appearance from highly concentrated and regular to
very chaotic and dispersed.  A further indication of their diversity of
properties is the order-of-magnitude spread in their masses and X-ray
fluxes (Smail et al.\ 1997b).  This is not to say that our small sample
spans the distribution of richness and types that are characteristic of
the Abell catalog, for example, but not all of the clusters are analogs
or richer versions of such unusual low-redshift clusters as Abell 1656
(the Coma cluster) or Abell 2029.

The images cover the central 0.4--0.8 h$^{-1}$ Mpc for a $q_o=0.5$ and
$h = H_o / 100$ km sec$^{-1}$ Mpc$^{-1}$ cosmology, for which 1 arcsec
is equivalent to 3.09 h$^{-1}$ kpc for our lowest redshift cluster and
3.76 h$^{-1}$ kpc in the most distant.  For a resolution approaching
0.1 arcsecond, then, these images show detail at the level of $\sim$500
pc, equivalent to observing galaxies in the Coma cluster with 1"
seeing.  While cruder than the resolution usually available for
morphological classification of nearby galaxies, this is sufficient for
the identification of basic morphological information --- the presence
of disks and bulges, and large-scale spiral arms, dust lanes, and bars.
In particular, this resolution is comparable to that obtained in the
D80 study of morphology in low-redshift clusters, for which z = 0.04
and 1" seeing were the norm.

Object identification and measurement of photometric parameters were
accomplished using a modified version of the SExtractor image analysis
package (\cite{ba96}).  Because of the low sky background and long
integration times, the exposures go very deep, to a 5$\sigma$ limiting
depth of $I_{814} \simeq 26.0$ or $R_{702} \simeq 27.0$.  However,
after considerable testing, including intercomparisons of our several
human classifiers, we concluded that reliable classification could not
be done fainter than  $R_{702} = 23.5$ or $I_{814} = 23.0$ with these
images.  In fact, as discussed below, for the purposes of the analysis
described here, we reached this limit only for the highest redshift
clusters.  In S97a there is a detailed discussion of how the
morphological classifications were assigned and what is their inferred
accuracy.  

For the purposes of this work, only the objects which were given
Revised Hubble types were used; galaxies that appeared extremely chaotic
or insufficiently detailed to fit into one of the categories
--- a few percent in all --- were not included.  This results in a
small inconsistency with what has been done in studies of lower redshift
clusters, for which galaxies were all assigned to one type or another,
but we believe it is the safest thing to do for the classifications of
distant samples.  It is certainly not the case that all of this few
percent of cases belong to one class, for example, the irregulars.
Probably half are cases of bulge-dominated, earlier types whose
division into elliptical, S0, and spiral cannot be made because of too
few pixels in an image.  Of the extremely chaotic systems, we believe
that most of these should not be put into the irregular class, since
they are much rarer at the present epoch and, as such, have no 
counterparts in the Revised Hubble scheme.  (The ``irregular'' class
in the Revised Hubble system refers to a specific morphological type,
and was not intended as a reposititory for galaxies that did not
fit into the other classifications.)

\section{The Morphology-Density Relation, Revisited}

Following Oemler (1974), Melnick and Sargent (1977) found a strong
gradient in morphological type with radial distance, what we will here
refer to as a T-R relation, for a sample of clusters with strong
thermal X-ray emission, in the sense of spiral fraction increasing
outwards and elliptical and S0 fractions increasing inwards.  As is
typical of luminous X-ray clusters, these were centrally concentrated
clusters with ``regular'' (non-clumpy) galaxy distributions.  The
DCAT80 sample of 55 clusters with $z \sim 0.04$ contained examples of
this type, all of which showed strong radial population gradients.  In
addition, however, this sample included many irregular clusters with
poorly defined centers and low concentration.  As Oemler had found,
radial gradients for these were weak or even absent; multiple centers
often complicated the attempt to look for such trends.  However,
because such clusters often were found to contain several higher
density clumps with increased fractions of E and S0 galaxies, Dressler
was led to look for a correlation of galaxy type with the local
density, as discussed in Appendix 1.

The basic interpretation of the \MD relation was that the environmental
density, possibly reaching back to quite early epochs, might have
exercised substantial influence on the evolution of morphological
types.  At the time this might have been considered surprising, since
it was commonly argued that galaxy mixing in the virializing collapse
of a cluster would remove any correlation of a galaxy with its previous
environment.  However, subsequent N-body simulations confirmed that simple
``top hat'' analytical models of cluster virialization were misleading
because the more realistic N-body models of structure growth showed
that mixing was far less thorough.  In fact, the simulations showed that galaxies found at the present epoch in high density regions were more than likely to have been in higher density regions throughout their lifetimes, 
even if there had been considerable mixing of the individual galaxies with
those from somewhat lower density environments (see, e.g., Evrard,
Silk, \& Szalay 1990).

The data from DCAT80 have been reanalyzed for the purposes of this
paper.  Following Whitmore, Gilmore, and Jones (1993), we have made a
field correction that is dependent on morphological type, in addition
to correcting the local surface density for field contamination.  The
percentages 18/23/59 adopted by these authors for field E/S0/S+I
contamination have been adopted here as well. This produces a small but
noticeable change --- in particular it lowers the spiral fraction at
the lowest projected densities, since many of these are in fact
projected field galaxies not associated with the cluster.  We have also
made finer divisions in binning the data, now 0.2 dex in surface
density, to make full use of the large, statistical sample.  This
revised version of the D80 morphology-density relation for the entire
55 cluster sample is shown in histogram form in Fig.~1.  Also shown in
Fig.~1, and in subsequent \MD plots, is the number histogram of the
membership in each bin.  This is useful for estimating when the
fractions in individual bins are very uncertain, i.e., when fewer
than 10 galaxies comprise a bin.  However, here we mean to compare only
overall trends and not individual bins, so we have chosen to omit error
bars for each bin, as they greatly confuse the diagrams.  (We have,
however, tabulated these values and their errors in tables as
Appendix 2.) The noise due to counting statistics can be judged by 
the scatter from one bin to the next of what would otherwise likely 
be monotonic relationships.

Subsequently, the \MD relation was studied for poorer groups and the
field (Bhavsar 1981, de Souza et al.\ 1983, Postman \& Geller 1984,
Giovanelli, Haynes, \& Chincarini 1986) with the general result that
the trends found for rich clusters were found to continue to lower
density environments, with lower precision due to the uncertainties of
projection effects for low-density environments.  However,
Salvador-Sole et al.\ (1989) pointed out that if the \MD relation were
universal for surface densities, as D80 had suggested, then the
relationship between morphology and three-dimensional space density
could not be universal, implying that a further dependence on some more
global property, one that compared whole clusters with one another,
might also play a role.  In fact, Dressler had suggested that his
sample showed some evidence for a global dependence on cluster
concentration by noting a shift of the population levels for the
clusters with the strongest X-ray emission.  Salvador-Sole et
al.\ pointed out that the sense of this effect, if real, was to imply
an even greater dependence on concentration in the three-dimensional
correlation.  Though this analysis might be questioned because of its
reliance on a smooth power-law model for the distribution of galaxies
in the cluster, both results seemed to point to some additional
determinant of cluster population beyond the local density, perhaps
connected with a global property of the cluster.

The case for a dependence on cluster populations with a global cluster
property was prosecuted even more vigorously by Whitmore \& Gilmore
(1991) and Whitmore, Gilmore, \& Jones, (1993), who argued that the
{\it sole} determinant of galaxy type within rich clusters is the
radial distance from the cluster center (a T-R relation) to the
exclusion of local density or even any other global property, such as
X-ray luminosity, velocity dispersion, or central density. A thorough
examination of this approach by Whitmore and collaborators is beyond
the scope of the present paper, but we include in this paper in
Appendix 1 a discussion of additional evidence that bears on this
question of whether local density or radial distance from the cluster
center is better correlated with morphological type.

Although we will show both \MD and T-R relations for our distant
clusters, our emphasis will be on the former.  In particular, we divide
our sample into centrally-concentrated, regular clusters and
low-concentration, irregular clusters.  A comparison of these two kinds
of clusters in the DCAT80 sample continues to provide the best evidence
that the \MD relation is a manifestation of a genuine physical process
operating in clusters that relates environmental density to galaxy
morphological type, either causally or through a common connection.  We
show such data in Fig.~2a for 10 of the most centrally concentrated and
Fig.~2b for the lowest-concentration clusters. The global properties of these
samples are very different: velocity dispersions and x-ray fluxes are
much higher for the centrally concentrated clusters, and of course,
cluster centers are well defined in these but not in the low
concentration clusters.  Nevertheless, the \MD relations for both kind
of clusters look very similar, with a strong dependence of spiral and
elliptical fraction with local density for both.  As we show in the
Appendix, the increase in elliptical fraction with $\Sigma$ is, within
the errors, the same in both, although there might be a significant
excess of S0's, and corresponding smaller spiral fraction, in the
centrally-concentrated clusters compared to the low-concentration
clusters.  This may be indicative of an additional factor other than
local density influencing cluster populations, but these differences
are marginally significant at best.

Because the \MD relation is evident for low-redshift clusters of all
types, we use it as the  basis for analyzing the \MD relation for
the intermediate-redshift clusters of our sample.  While global
influences may also play a role for low-redshift clusters, the evidence
continues to support the conclusion that local density and morphological
type are strongly connected.

\newpage
\section{The \MD Relation at $z \sim$ 0.5}

We now turn to the subject of how the \MD or T-R relations are manifest
in our sample of 10 intermediate-redshift clusters at $z \sim 0.5$.  A
list of these clusters, their positions, and redshifts are given in Table
1. We will assume that our morphological classifications are made
comparably to the DCAT80 study of 55 nearby clusters, that is, that the
ability to discern salient features for classification are not
appreciably diminished for these more distant clusters.  With the
possible exception of a difficulty in distinguishing face-on S0 from
elliptical galaxies, a problem even for nearby galaxies, we believe
that this assumption is justified, as discussed in S97a.  (We also
discuss this issue further below.) As in DCAT80, our classifications of morphological types in the intermediate-redshift clusters include cases designated E/S0 and S0/E. As explained in S97a, the order reflects the 
preference in the classification.  Therefore, as done in that study, we 
have added these types into E and S0, respectively.

As in D80, the analysis of the \MD relation for the DCAT80 sample was
done for galaxies down to a fixed apparent magnitude limit $V = 16.5$,
which corresponds to $M_V = -20.4$ for $H_o$ = 50 and z = 0.040.
We have chosen $M_V = -20.0$ as the magnitude limit of the \MD analysis
of our clusters at intermediate redshift. Using standard
Kron-Cousins/Johnson filter transmissions, the energy distribution of
an Sbc galaxy, and conversions from F702W to R and F814W to I given by
Holtzman et al. (1995), we have derived the magnitude limits given in
Table 1.  Also in Table 1 we give the number of galaxies in each
cluster brighter than the magnitude limit, and the fractional
representation of the different morphological types for each field, 
corrected for field contamination.  We obtained values of Galactic
extinction for each cluster from NED; the maximum of these corresponds
to 0.23 mag in the F702W band observation of CL0303+17.  These small
corrections to the limiting magnitude have not been made, as also was
the choice in D80.  They make no significant change to the fractional distributions of the populations.  

We also need to take note of the fact that our intermediate-redshift
sample is studied over a smaller area (by a linear factor of
approximately 2.5) than that of the low-redshift sample of DCAT80.  We
have therefore reanalyzed the data for the 55 clusters for a restricted
area --- a box that is 1094 arcsec on a side, or 1.2 Mpc ($H_o$ = 50,
adopted in D80), for $z = 0.040$, and scales with redshift. For most of
the low redshift clusters we have centered the box on the plate center,
but have used the ``local density peak'' centers (see Appendix 1) for
the centrally concentrated and low concentrated subsamples.  The \MD
relation for the whole sample, and for the centrally concentrated and
low concentration samples, are shown in Fig.~3a-c, respectively. There
is no obvious difference between these and Figs.~1 \&~2, except that
the lowest density range is unpopulated, as expected, and that the
statistics for the two subsamples are substantially degraded,
especially for the low concentration sample.

We have applied a field correction, taken from the Medium Deep Survey
(\cite{grif94}), of 11.2 gal/sq arcmin at $I = 23.0$.   As we did for
the 55 cluster sample, we have made differential corrections by type to
the counts in the distant clusters.  We adopt the percentages of the
different types found in the Medium Deep Survey --- 10\%, 10\%, and
80\% for elliptical, S0, and spiral galaxies, respectively (S97a).
These values are, coincidentally, the same as used by D80, which have
been replaced here in our reanalysis of the DCAT80 sample with the
Whitmore, Gilmore and Jones's values of 18\%, 23\%, and 59\%).  There
is no reason or expectation for using the same values for both low- and
intermediate-redshift clusters, since the resolution of the images, and
the composition of the field contamination are not necessarily the same.

We now compare Fig.~3a (or, equivalently, Fig.~1) to the \MD
relationship for the entire $z \sim 0.5$ sample, Fig.~4, noting first
that, while there is considerable overlap in density between the two
samples, the density range encompassed by the more distant sample is
shifted by half a dex to higher density.  This undoubtedly reflects the
fact that these clusters are systematically richer than the {\it
typical} clusters of DCAT80, although there is considerable overlap
with the projected density range covered by that sample. Specifically,
13\% of galaxies in the DCAT80 sample lie at $\Sigma < 1.2$ lower limit
of the intermediate redshift sample, and 12\% of this sample are found
at $\Sigma > 2.5$, a higher projected density than found in any of the
DCAT80 clusters.  Because 75\% of the galaxies in the two samples share
the same range in projected density range (somewhat greater if $q_o <<
0.5)$, we do not consider these two samples as composed of qualitatively
different kinds of clusters.  (For example, CL1601, CL0054, and Cl0303,
share the density range of the DCAT80 sample.)  It would be desirable,
of course, to include more clusters typical of the present epoch in
future samples.

Before addressing the question of gradients in Fig.~4, we take note of
the obvious differences between this and the nearby cluster sample.  As
is now well known, spirals are greatly overabundant in high
density environments compared to present-epoch clusters, but, perhaps
surprisingly, the difference seems to be made up entirely by a paucity
of S0 galaxies rather than an underabundance of {\it both} S0 and E
galaxies.  In fact, E galaxies appear to be in even greater abundance
than in the nearby sample!  In the overlapping density regimes of
Fig.~3a and Fig.~4, spirals are a factor of 2 overabundant, S0's are a
factor of 2--3 underabundant, and ellipticals are a factor of 1.5
overabundant.  The E or S0 population is, of course, little affected
by the uncertainty in the correction for field galaxies, since galaxies 
of this type make up a small ($\sim20\%$) fraction of the 
intermediate-redshift field population. 

In order to further investigate the possibility of misclassifications
of E and S0 galaxies we have, as explained in S97a, compared the
distribution in flattenings of S0 galaxies in our sample with that for
the Coma cluster, in an attempt to see if we have misclassified S0
galaxies as ellipticals, particularly for the face-on cases.  The good
agreement of these distributions indicates that this is not the case.
Our best estimate is that we have misclassified as ellipticals
approximately $12\%$ of the total S0 population that is nearly face
on.  As discussed in S97a, this has been determined through comparison
with the Revised Shapely Ames galaxies (as tabulated in Sandage,
Freeman, and Stokes 1970), but we believe that in comparison with
classifications of nearby clusters, such as DCAT80, the degree of
misclassification is so similar as to make any systematic differences
negligible.  We demonstrate this in Fig.~5, where we compare the
distribution of ellipticities for both E and S0 in the $z \sim 0.5$
cluster sample with the E and S0 galaxies in the Coma cluster (Andreon
et al.\ 1996) and 11 clusters from DCAT80 with $0.035 < z < 0.044$. In
both cases the distributions are in very good agreement, suggesting
that any {\it differential} comparison of the E and S0 fractions in
these clusters is justified.  We thus conclude that the substantial
difference seen in the S0/E fraction in distant clusters compared to
nearby clusters is genuine for this sample; of course, verification in
a larger sample, to be certain that our sample is truly representative, 
would  be highly desirable.  The implications of this difference in 
E and S0 populations is discussed in Section 5.

We now ask whether any trend of morphology with density is apparent for
the distant sample.  From Fig.~4 it appears that a modest \MD relation
is present, but it is only for the bins of highest density --- over the
last factor of 5 in surface density.  Over this range the spiral
fraction plummets and the elliptical fraction rises sharply, but for
the lower density zones, over which there is a very noticeable gradient
in the nearby clusters, the relationships are basically flat.  Slicing
the sample by redshift or richness does not change this result.
However, when the sample is divided by the cluster concentration, which
correlates well with degree of regularity, a very different picture
emerges. We have used Butcher \& Oemler's (1978) definition of
concentration as $\log (R_{60}/R_{20})$, where the latter refer
to the radii containing 60\% and 20\% of the cluster populations,
respectively.  These values are given in Table 1 under the heading 
``Conc.''  Fig.~6 shows the \MD relation for the four clusters of
the $z \sim 0.5$ sample with the highest central concentration ---
3C295, Cl0024+16, Cl0016+16, and Cl0054$-$27.  The \MD relation for
this subset is steep and well defined over the entire density range, as
strong as the \MD relation for the low redshift sample.  As expected,
the T-R relation (Fig.~7), being more-or-less degenerate with \MD for
these cases, is also very strong for these clusters.  Though the
gradients in Fig.~6 are as strong as that found for low-redshift
clusters, there are substantial differences, of course.  Most obvious
is the prevalence of spiral galaxies even in very high density
regions.  And although the S0 galaxies show the same flat distribution
in projected density, they are, as we have pointed out, far less
abundant.  Finally, in comparison with Fig.~3, Fig.~6 appears to show a
5--10\% excess of elliptical galaxies at the same surface density,
extending to much higher fractions, at higher density, than has been
found for lower redshift clusters.  This may again highlight the
unusual richness of some of the high redshift clusters in this sample.

The remaining clusters of lower concentration --- Cl0303+17,
Cl0412$-$65, Cl0939+47, Cl1447+23, Cl1601+42 --- have irregular, clumpy
distributions of galaxies, just like their low redshift counterparts.
We show in Fig.~8 that, in contrast with the relatively strong
gradients seen for the centrally concentrated clusters, there are no
correlations at all for these 5 lowest-concentration, irregular
clusters, except perhaps for a small {\it anticorrelation} of low
statistical significance.\footnote{We have omitted the outer regions of
two clusters, A370 and Cl0939+47, cataloged in S96.  The central region
of Cl0939+47, a low concentration cluster, is included in our analysis
here, but equivalent morphological classifications using WFPC-2 for the
center field of A370, a centrally concentrated cluster, are not yet
available.} (The T-R relation is, of course, nonexistent as well.)
Apparently, then, the morphology-density relation is well established
at higher redshift for the clusters that appear well evolved
dynamically, but there is no apparent segregation in clusters that have
more chaotic distributions.  Of course, even the equivalent diagram for
the centers of low-z clusters of low concentration (Fig.~3c) suffers
from small number statistics, so it is possible that this is an
unrepresentative sample.  We note however, that although the population
gradients are not evident in the high-z, low-concentration clusters,
the elliptical fraction is high and the S0 fraction is very low, as
it is for the high-z, high-concentration clusters and the low-z sample.
It is only the arrangement of these galaxies within the clusters, and
not their relative proportions, that seems to differ from the centrally concentrated clusters.

These, then, are our principal results from the small sample of
intermediate redshift clusters we have studied. (1) Over the range of
density studied, which extends to somewhat higher densities, on
average, than for the typical nearby cluster, elliptical galaxies are
present in numbers comparable to what is seen today, while spirals are
much more abundant and S0 galaxies much less abundant. (2) There is a
modest \MD relation, but only for the highest densities, when all the
clusters are taken together.  (3) When the sample is divided into 
high-concentration, regular clusters and low-concentration, irregular
clusters, the former show strong gradients in morphology over the
represented range in surface density, but the latter show no gradients
at all.

\section{Discussion}

Perhaps the most striking result of this study is that, whether the
clusters appear dynamically ``mature'' or not, the incidence of
elliptical galaxies is already very high.  Put another way,
the elliptical fraction is high whether or not ellipticals are collected 
into dense, central regions.  This suggests that elliptical galaxies 
predate, and are basically independent of, the virialization of a 
rich cluster.  

Supporting evidence for the idea that elliptical galaxies in rich
clusters are old can be found in the Ellis et al.\ (1996) study of the
$(U-V)_o$ colors of these same ellipticals.  Ellis et al. concluded,
based on a scatter in rest frame $U-V$ color of 0.07 mag --- scarcely
larger than the value of 0.035 mag found for ellipticals in the nearby
Coma cluster (Bower, Lucey, \& Ellis 1992) --- that the stars in
cluster elliptical galaxies formed early, $z \gtrsim 3$ (or that, if
they formed later, they formed within a very small time).  This seems
to argue against a ``late-merger'' origin for ellipticals, or at the
least against dissipative mergers of star forming systems.  In fact, as
we will explore elsewhere, those systems in our sample for which a
merger is likely to have ``recently'' occurred are still, for the most
part, disk systems; therefore, most are unlikely to be the ancestors of
elliptical galaxies because there is no expectation that the disk will
be destroyed without another significant encounter.  We have further
evidence of this in their distribution within the clusters: when we
plot the fractional population of highly asymmetric or disturbed
morphologies (all morphological types with $D > 1$, as defined in S97a)
with local surface density (Fig.~9), we see that these follow the S0 or
spiral trend rather than the trend for ellipticals.\footnote{The
density range extends to lower values than previously because no
attempt has been made at a field correction.}  That is, the fraction of
asymmetric galaxies is decreasing, or perhaps flat, but certainly not
rising, with increasing density.  This is a result similar to that
found by Oemler, Dressler, and Butcher (1997) for the distribution of
``disturbed'' galaxies in four distant clusters.

Taken together, these three observations --- the prevalence of E
galaxies, their small $U-V$ scatter, and the fact that most of the
disturbed galaxies are disk systems and likely to remain so --- suggest
that, at least for the environments of rich clusters, elliptical
galaxies are not the result of mergers of star-forming, gas-rich
systems after a redshift $z \sim 3$.  This does not preclude the
possibility of dissipationless mergers at $z \sim 1$, say, when these
clusters might have been in the process of accreting small groups with
lower velocity dispersion, but both the distribution and numbers of
ellipticals we have found here, and their photometric and spectral
properties, suggest that ellipticals in these regions are not produced
by late mergers, or in any process that depended on the dynamical
evolution of a rich cluster.  Instead, gaseous mergers or coherent
collapse at high redshift, or growth of spheroids through
dissipationless mergers continuing until later epochs, seems to be the
history indicated for ellipticals.  It is remarkable, we think, that
the environments of proto-clusters of this richness were able to
produce such a large population of ellipticals before the identities of
the clusters themselves were well established.

The situation for the S0 galaxies seems to be just the opposite.
Though the ones we find are, like the ellipticals, red and with little
scatter in color, their numbers are so deficient as to suggest that
many need to be added since $z \sim 0.5$, in order to reach the
populations of present-epoch clusters.  The source of these S0's seems
clear: the overabundance of spirals provides a reservoir of galaxies
which may be stripped by ram pressure, tidally harassed (Moore et
al.\ 1995) merged, or subjected to strong 2-body gravitational
interactions, with the result of producing today's dormant disk
galaxies in clusters.  As we will describe in more detail in a later
paper, our $z \sim 0.5$ cluster sample includes a significant number of
disturbed, distorted morphologies, often with spectroscopic evidence of
strong episodes of recent --- in a few cases current --- star
formation.  These may be the result of mergers, strong interactions,
accretions, harassment, or stripping --- we are still unable to tell
which of these processes are responsible.  But, as we have said, we do
know from our morphological classifications that most of these are disk
systems --- they do not seem destined to settle into luminous
ellipticals galaxies when their jostling and bursts of star formation
have ceased.  Though the exact mechanism(s) may be yet unspecified, it
seems that at least half of the S0 galaxies in today's clusters have
been made by such processes since $z \sim 0.5$.  Some support for this
model may also be found from the trend of the S0/E ratio with redshift,
which we show in Fig.~10.  Omitting the two outer fields, which are
properly excluded for this analysis, we see what appears to be a
significant trend of increasing S0 abundance with decreasing redshift.
Even more provocative is the observation that this trend, in linear
extrapolation, comes close to predicting the S0/E ratio of $\sim$2
found in the D80 study of clusters at low redshift. Although these data
are of too low statistical weight to be conclusive, they suggest that
we could be seeing the process of S0 production over the interval in
cosmic time probed by these observations.

On the other hand, according to Ellis et al.\ (1996), the spread in rest
frame $U-V$ colors for S0 galaxies in these clusters is as small as it
is for ellipticals, from which it was concluded that the stellar
populations in ellipticals are very old.  While we may with the present
data simply postulate that these represent a population of old S0's
that formed with the ellipticals, our suggestion that most of the
S0's in these clusters have yet to form means that the color
spread at some later time, at $z \sim 0.2$, for example, should be
significantly larger, reflecting the more recent evolution of these
systems from previously star forming galaxies.  To make comparable
morphological classifications in clusters at this redshift will require
either a mosaic coverage with HST or larger-field CCD imaging with
ground based images under conditions of superb seeing.

At first glance it may not be surprising, then, to see a \MD relation
for the dynamically well-evolved clusters of our sample.  One can
imagine a large initial population of elliptical galaxies, formed at
some earlier time, being diluted by infalling disk galaxies in
increasing numbers working out from the cluster center.  Any mechanism
for converting spiral galaxies to S0's that is more efficient with
increasing density would account qualitatively for the \MD relation
observed at $z \sim 0.5$ and $z \sim 0$.  In this connection it is
interesting to note that in present-epoch low-concentration, irregular
clusters (Fig. 3) ellipticals appear to outnumber S0's in the highest
density regimes, whereas the opposite situation seems to hold for the
centrally concentrated clusters.  Again, the manufacture of S0's by
some density-dependent process seems to be a reasonable
interpretation.

What prevents this picture from being a tidy one, however, is the fact
that the 5 low-concentration, more irregular clusters in our distant
sample show no concentration of ellipticals towards the densest
regions.  Instead, they possess large numbers of ellipticals spread
willy-nilly over the area surveyed.  Does the gradient we see in the
regular clusters mean that dilution alone by infalling galaxies has
already been sufficient to set up an apparent concentration of
ellipticals, or must there also be a mechanism for bringing ellipticals
to what will become the dense cluster regions, and, as the Whitmore \&
Gilmore result suggests, even more strongly to one specific cluster
center?  Dynamical friction is the obvious candidate for concentrating
the ellipticals as different subgroups merge, but would it be
sufficiently effective?  By virtue of their luminosity function, which
has the brightest characteristic magnitude of the morphological types
(Sandage, Binggeli, \& Tammann 1985), and their high M/L ratios (Faber
\& Gallagher 1976), ellipticals are, on average, significantly more
massive than most cluster galaxies, but it remains to be seen if the
effect is sufficient to generate a substantial segregation.  On the
other hand, the lack of a radial gradient in the luminosity function in
clusters of galaxies argues against this effect being an strong one.

This discussion begs what must be the more puzzling question: why are
the irregular clusters at $z \sim 0.5$ different in terms of their \MD
relations from similar clusters today?  Are any of these distant
irregular clusters destined to remain so until the present epoch, in
which case, mechanisms like dynamical friction and dilution through
infall are certainly needed to generate trends that were not present at
$z \sim 0.5$?  Alternatively, are these distant irregular clusters the
ancestors of some of today's regular clusters? If so, the same
processes must have occurred to create the \MD relation, but there must be
an {\it additional} explanation of why irregular clusters today, which
are almost certainly younger dynamically than today's regular clusters,
are amalgams of groups that already have, individually, well
established segregations of galaxy types.  A reasonable explanation for
this behavior might be that morphological segregation has occurred
hierarchically over time, first for the densest, most populous groups,
and then working its way down to the smaller, less dense systems.  In
this picture, the groups that make up the irregular clusters at $z \sim
0.5$ had not yet undergone significant morphological segregation, but
by the present epoch, even these groups, from which the Hercules
cluster (Abell 2151) is now being composed, would exhibit such
correlations.  Observations of smaller groups at intermediate redshift
will be crucial for testing this notion.

Of course, testing these schematic ideas depends on understanding which
mechanisms are responsible for the manufacture and distribution of S0
galaxies and, at an earlier epoch (we believe), the ellipticals.  Our
sample is still too small to say with certainty that these trends are
universal; a larger sample is essential to verify what we have found
here.  Furthermore, our present data do not include sufficient area
from the outer regions of intermediate-redshift clusters, a comparison
with which might allow for a discrimination between the various
candidates for S0 production.  In forthcoming papers we hope to throw
some light on this issue through the application of our large sample of
spectroscopic data and what it tells us about the history of star
formation in these distant cluster galaxies.

It is clearly also the case that a push to higher redshift could be
very revealing: it would seem probable, based on what we have found,
that the few clusters that exist at $z \sim 1$ will show little
morphological segregation of any kind.  Also,  by filling in the
evolutionary sequence of clusters at $z \sim 0.2$ with morphological
types and colors, the progression of cluster evolution, particularly
for the irregular clusters, might be better understood.

\section{Conclusions}

We have revisited the relation between morphology and surface density
and confirmed that it is a fundamental characteristic of all nearby
clusters and groups.  However, for a sample of distant clusters, $z
\sim 0.5$, studied with HST, the \MD relation is apparent only in high
concentration, regular clusters that are presumably dynamically
evolved.   No correlation is found for less evolved, low concentration,
irregular clusters.  This suggests that the mechanisms that produce
morphological segregation, either directly, by dynamical friction, for
example, or through converting one type of galaxy into another, may
work at different rates depending on the mass scale of the group or
cluster.  By the present epoch, even the smaller groups, presently
coalescing to form today's irregular clusters, may have accomplished a
morphological segregation that was absent only 1/3 of a Hubble time ago.

The large number of elliptical galaxies in these clusters, $\sim$40\%
in both regular and irregular clusters, suggests that the creation of
ellipticals predates cluster virialization.  If mergers are responsible
for making the ellipticals that now inhabit these rich clusters, they
must have been dissipationless, in the ``group phase'' at $z \sim 1$,
or much earlier, $z > 3$, if significant dissipation and star
formation were involved.  In contrast, the relative paucity of S0's in
the intermediate redshift clusters suggests that many of them have
indeed been added since $z \sim 0.5$, by mechanisms that acted on the
excessive numbers, compared to today's clusters, of spirals and
irregulars.
  
\section*{Acknowledgements} Firstly, we wish to thank Ray Lucas at
STScI for his enthusiastic, able help in the efficient acquisition
of these HST observations. AD and AO acknowledge support from NASA
through STScI grant 3857.  IRS, RSE and RMS acknowledge support from
the Particle Physics and Astronomy Research Council.  WJC acknowledges
support from the Australian Department of Industry, Science and
Technology, the Australian Research Council and Sun Microsystems.
This work was supported in part by the Formation and Evolution of
Galaxies network set up by the European Commission under contract ERB 
FMRX-CT96-086 of its TMR program.

\newpage
\section*{Appendix 1: A Further Look at the \MD Relation for Nearby
Clusters}

The \MD relation, the basis for our analysis in this paper, has been
challenged as the primary correlation between galaxy type and some
associated property of the galaxy clusters.  Our purpose here is review
briefly the derivations of this relationship and to comment on these
objections.

A first step is to reexamine the raw data that led Dressler to suspect
that galaxy populations are in fact correlated with local density.  In
Fig.~11 we show representative clusters from the samples, with the E
and S0 galaxies highlighted.  Although only qualitative, one can
observe even in these maps that the E and S0 galaxies are clumped
together much more strongly than the cluster spirals, which are spread
almost uniformly over the fields.  This is true not only for reasonably
regular clusters, such as Abell 151, but even for clusters without a
well defined, single high density region, such as Abell1631, DC0003-50,
and DC0326-53, all of which are shown in Fig.~11.  Although it is
difficult to use \MD relations to compare individual cases, we can
verify the subjective impression of the greater clumping of E and S0
galaxies by a simpler statistic, the average nearest neighbor distances
for the different types.  We divided the sample into two types, E+S0
galaxies, and spirals.  For the clusters just listed, the mean distance
(in h$^{-1}$ Mpc) from an E or S0 galaxy to its nearest-neighbor of
the same type, compared to the mean distance of ``nearest-neighbor spiral galaxies'' are 0.202:0.344, 0.394:624, 0.365:0.459, and 0.338:0.551, respectively.  In each case the nearest-neighbor distance between spirals 
is substantially larger.

This result is consistent with the idea, unsubstantiated in 1980 but
subsequently strongly supported by radial velocity studies, that
subclustering is a real and important phenomenon is most clusters.
While the change in galaxy populations with radial distance from the
centers of the clusters can be seen in Fig.~2, the existence of
subgroups with populations shifted towards early Hubble types, even
when removed from the centers, is also apparent.

Whitmore and collaborators claim that the T-R relation is more
fundamental than the \MD relation in understanding the distribution of
galaxy types within a cluster.  In order to evaluate this claim, it is
important to remember that the \MD and T-R relations are nearly
degenerate for those clusters with a reasonably smooth, regular
distribution of cluster galaxies.  In such cases local surface density
correlates very well with clustocentric radius. Since
the remaining variations are statistical and not representative of
substantial subclustering, there is no reason to expect to be able to
to distinguish between the two candidates for a fundamental
relationship.  Thus, it is essential to compare these relationships for
both centrally concentrated regular clusters {\it and} clumpy,
irregular clusters to compare the efficacy of local density or distance
from the cluster center as a driver of galaxy type.

Such a comparison was the basis of Dressler's claim that the \MD
relation is universal, in the sense that it held irrespective of other
global parameters.  For example, Dressler noted that the \MD relations
for the high-concentration and low concentration clusters were very
similar, perhaps indistinguishable within sampling errors.  In order
to make quantitative comparisons between the relations of Fig.~2, we have
fit the slopes and intercepts of the elliptical and spiral fractions as
a function of $\Sigma$. For Figs. 2a and 2b respectively we find that
the f(E) = 0.061(0.069) + 0.122(0.026)$\Sigma$ and f(E) = 0.094(0.041)
+ 0.090(0.039)$\Sigma$ (with 1 $\sigma$ errors given in parentheses),
thus confirming the impression that these cannot be distinguished
within the errors.  For the spirals, f(Sp) = 0.623(0.044) -
0.250(0.030)$\Sigma$ for Fig. 2a and f(Sp) = 0.642(0.069) -
0.184(0.058)$\Sigma$ for Fig. 2b, which indicates a marginally
significant difference in the sense of a lower fraction of spirals, and
a higher fraction of elliptical and S0 galaxies at a given density for
the high-concentration sample.  This difference might suggest a
dependence on the cluster type, a global parameter, in addition to the
\MD relation, but the evidence is weak.

While the \MD relation holds to first order for both regular and
irregular clusters, the T-R relation explored by Whitmore and
collaborators shows very different behavior in the two types of
clusters.  In Fig.~12 we show the T-R relation for the same samples of
Fig.~2, using center positions defined by the peak in local galaxy
density over the field (values taken from Beers and Tonry 1986).   As
found by those studies, the T-R relation for the centrally concentrated
clusters is strong.  In contrast, the relationship for irregular
clusters of low concentration is weak.  In fact, it is arguably absent
except for the points within 0.5 Mpc, which result from the peaking up
on a local density enhancement, in which domain one would expect the
\MD relation to be again apparent, which it is.  Our interpretation of
Fig.~12 is, then, that application of the T-R has weakened or even
eliminated a true \MD dependence further out in the cluster, by angular
smoothing within the annular rings out what genuine variations do
exist, leaving only the sharp gradients in the central region.
Whitmore and collaborators have used the flat-to-steep profile to argue
that the T-R relation is the fundamental one, but our analysis here
argues, as did Dressler, that the dependence on distance from a single
cluster center is only a second-order, though likely important
supplement to the more fundamental correlation with local density.

That fact that the \MD relation holds for both types of clusters, and
the T-R relation is substantially different for the two types, remains
the best evidence that it local density is fundamental to the 
determination of galaxy type.

A further challenge has been made by Sanroma \& Salvador-Sole (1990).
These authors used angular scrambling of the galaxy positions in DCAT80
to argue that that, for the entire sample, the \MD does not exist, but
is instead a derivative of the T-R relation.  Again, separating
centrally concentrated from low-concentration clusters is important.
We have repeated this experiment for the samples of Fig.~2, scrambling
around both "density peak" and "median" centers.  The ranges in the
distributions in density are largely maintained by the scrambling
procedure, so a Kolmogorov-Smirnov test can be applied.  We find that
scrambling the centrally concentrated clusters produces no significant
change in the \MD relation: the fraction of E galaxies at a given
projected density for the real data would match that of the scrambled
data 38\% of the time, and the spiral fraction would match 18\% of the
time.  In stark contrast, the \MD relation for the low-concentration
clusters is in fact destroyed by the scrambling: all probabilities are
$<10^{-4}$, regardless of choice of galaxy center.

It is our contention, then, that the \MD relation remains important for
understanding the distribution of morphological types in present-epoch
clusters.  This is not to say that there are no global dependences of
cluster population, as Whitmore and collaborators, Sanroma \&
Salvador-Sole, Oemler, and even Dressler himself have suggested.  In
fact, in addition to the evidence mentioned above, both Whitmore \&
Gilmore's T-R  relation, and the scrambling analysis of Salvadore-Sole,
provide substantive evidence for global components.  As can be seen in
Fig. 5 of Whitmore \& Gilmore (1991), the percentage of elliptical
galaxies rises even higher, with a compensating {\it drop} in the S0
population, for those galaxies closest to the cluster center as defined
by the presence of a ``D'' galaxy, compared to the bin of highest
density of the \MD relation.\footnote{One might have suspected that
increasing the resolution of the \MD relation, by decreasing the
``kernel'' size to a smaller number of neighbors, would have revealed
this same effect for the densest regions.  However, our tests indicate
that reducing to 5 neighbors does not significantly alter the
fractions.} The same can perhaps be inferred from our experiments of
scrambling the clusters of the high-concentration sample and comparing
the \MD relation to that of the real data. It appears that when all but
the centrally located high-density peak are eliminated by the
scrambling, the E fraction rises and the S0 fraction falls, as if the
central location of these galaxies has an additional effect on the
population.  So, it appears that central location may indeed influence
a galaxy's morphology, and be perhaps an important ``second-parameter''
to the \MD relation.

\vspace{8mm}
\section*{Appendix 2: Tabulations of \MD relations}

In Table 2 we tabulate the values of the \MD relations in the figures,
as well as their errors.  The errors stated in Table 2 are 1 sigma
errors derived from Poisson statistics, assuming that the number of
galaxies of a given type is independent of the number of galaxies of
other types. It is also assumed that the number of background galaxies
has a Poisson distribution (i.e.  that the angular correlation function
of galaxies is negligible at these magnitudes.)

\cleardoublepage

\centerline{\sc \hfil Table 1 \hfil }
 
\centerline{\sc \hfil Cluster sample and properties \hfil }

\begin{center} 
{\scriptsize
\begin{tabular}{lcccccccc}
\noalign{\medskip}
\hline\hline
\noalign{\smallskip}
{Cluster} & RA(J2000)  DEC(J2000) & {$z$} & Filter & M$_{\rm lim}$ & Ngal &  \% E:S0:Sp & Conc.\ \cr
\noalign{\smallskip}
\noalign{\hrule}
\noalign{\smallskip}
A370\#2      & 02:40:01.1  -01:36:45 & 0.37 & F814W & 21.98 &  71  & 28:20:52  &   ---- \cr
Cl1446+26    & 14:49:28.2  +26:07:57 & 0.37 & F702W & 22.24 & 107  & 31:26:43  &   0.30 \cr
Cl0024+16    & 00:26:35.6  +17:09:43 & 0.39 & F814W & 22.11 & 170  & 37:24:39  &   0.53 \cr
Cl0939+47    & 09:43:02.6  +46:58:57 & 0.41 & F702W & 22.48 & 124  & 31:24:45  &   0.34 \cr
Cl0939+47\#2 & 09:43:02.5  +46:56:07 & 0.41 & F814W & 22.23 &  72  & 35:07:58  &   ---- \cr
Cl0303+17    & 03:06:15.9  +17:19:17 & 0.42 & F702W & 22.54 &  93  & 39:19:42  &    0.31 \cr
3C295        & 14:11:19.5  +52:12:21 & 0.46 & F702W & 22.76 &  87  & 51:21:28  &   0.58 \cr
Cl0412$-$65  & 04:12:51.7  -65:50:17 & 0.51 & F814W & 22.77 &  91  & 47:07:46  & 0.30 \cr
Cl1601+42    & 16:03:10.6  +42:45:35 & 0.54 & F702W & 23.20 & 100  & 33:12:55  & 0.34 \cr
Cl0016+16    & 00:18:33.6  +16:25:46 & 0.55 & F814W & 22.96 & 193  & 58:15:27  &  0.49 \cr
Cl0054$-$27  & 00:56:54.6  -27:40:31 & 0.56 & F814W & 23.00 & 119  & 34:18:48  & 0.38 \cr
\end{tabular}
}
\end{center}

\vspace{4mm}
\centerline{\sc \hfil Table 2 \hfil }
\centerline{\sc \hfil Data for Figure 1 \hfil }
\begin{center} 
\begin{tabular}{lcccccc}
\noalign{\smallskip}
\hline\hline
\noalign{\smallskip}
 $\Sigma$  &  f(E)~~~$\sigma$  &  f(S0)~~~$\sigma$ &  f(Sp)~~~$\sigma$ & \ \cr
\noalign{\smallskip}
\noalign{\hrule}
\noalign{\smallskip}
 -0.05  & 0.000  0.000  &  0.460  0.148  &  0.540  0.148  \cr
  0.15  & 0.110  0.039  &  0.289  0.053  &  0.601  0.059  \cr
  0.35  & 0.090  0.021  &  0.320  0.032  &  0.589  0.034  \cr
  0.55  & 0.119  0.016  &  0.346  0.023  &  0.535  0.025  \cr
  0.75  & 0.133  0.014  &  0.348  0.020  &  0.519  0.021  \cr
  0.95  & 0.177  0.014  &  0.391  0.018  &  0.432  0.019  \cr
  1.15  & 0.182  0.014  &  0.425  0.018  &  0.392  0.018  \cr
  1.35  & 0.196  0.015  &  0.443  0.019  &  0.361  0.018  \cr
  1.55  & 0.225  0.018  &  0.473  0.021  &  0.302  0.020  \cr
  1.75  & 0.287  0.024  &  0.454  0.026  &  0.259  0.023  \cr
  1.95  & 0.376  0.033  &  0.448  0.034  &  0.176  0.026  \cr
  2.15  & 0.398  0.050  &  0.521  0.051  &  0.081  0.028  \cr
  2.35  & 0.360  0.096  &  0.561  0.100  &  0.079  0.055  \cr
  2.55  & 0.500  0.134  &  0.429  0.132  &  0.071  0.069  \cr
\end{tabular}
\end{center}

\newpage
\centerline{\sc \hfil Data for Figure 2a \hfil }
\begin{center} 
\begin{tabular}{lcccccc}
\noalign{\smallskip}
\hline\hline
\noalign{\smallskip}
 $\Sigma$  &  f(E)~~~$\sigma$  &  f(S0)~~~$\sigma$ &  f(Sp)~~~$\sigma$ & \ \cr
\noalign{\smallskip}
\noalign{\hrule}
\noalign{\smallskip}
 -0.05  & 0.000  0.000  &  0.589  0.250  &  0.411  0.250  \cr
  0.15  & 0.105  0.065  &  0.241  0.088  &  0.654  0.099  \cr
  0.35  & 0.106  0.040  &  0.362  0.061  &  0.532  0.064  \cr
  0.55  & 0.121  0.034  &  0.386  0.050  &  0.493  0.052  \cr
  0.75  & 0.155  0.036  &  0.379  0.048  &  0.466  0.050  \cr
  0.95  & 0.215  0.038  &  0.412  0.045  &  0.373  0.045  \cr
  1.15  & 0.156  0.035  &  0.498  0.049  &  0.345  0.047  \cr
  1.35  & 0.208  0.043  &  0.524  0.053  &  0.269  0.048  \cr
  1.55  & 0.201  0.042  &  0.563  0.052  &  0.236  0.045  \cr
  1.75  & 0.331  0.046  &  0.506  0.048  &  0.163  0.036  \cr
  1.95  & 0.435  0.069  &  0.416  0.068  &  0.149  0.050  \cr
  2.15  & 0.305  0.096  &  0.610  0.102  &  0.085  0.059  \cr
  2.35  & 0.364  0.146  &  0.547  0.151  &  0.089  0.088  \cr
\end{tabular}
\end{center}

\centerline{\sc \hfil Data for Figure 2b \hfil }
\begin{center} 
\begin{tabular}{lcccccc}
\noalign{\smallskip}
\hline\hline
\noalign{\smallskip}
 $\Sigma$  &  f(E)~~~$\sigma$  &  f(S0)~~~$\sigma$ &  f(Sp)~~~$\sigma$ & \ \cr
\noalign{\smallskip}
\noalign{\hrule}
\noalign{\smallskip}
 -0.05  & 0.149  0.242  &  0.595  0.328  &  0.256  0.316  \cr
  0.15  & 0.177  0.112  &  0.234  0.135  &  0.589  0.153  \cr
  0.35  & 0.113  0.066  &  0.192  0.079  &  0.695  0.093  \cr
  0.55  & 0.153  0.049  &  0.318  0.062  &  0.529  0.067  \cr
  0.75  & 0.169  0.038  &  0.298  0.047  &  0.533  0.051  \cr
  0.95  & 0.181  0.036  &  0.364  0.046  &  0.454  0.048  \cr
  1.15  & 0.180  0.031  &  0.457  0.040  &  0.363  0.039  \cr
  1.35  & 0.189  0.034  &  0.459  0.043  &  0.351  0.042  \cr
  1.55  & 0.252  0.055  &  0.393  0.062  &  0.355  0.061  \cr
  1.75  & 0.318  0.108  &  0.369  0.112  &  0.313  0.108  \cr
  1.95  & 0.481  0.101  &  0.280  0.090  &  0.238  0.086  \cr
  2.15  & 0.501  0.251  &  0.499  0.251  &  0.000  0.000  \cr
\end{tabular}
\end{center}

\newpage

\centerline{\sc \hfil Data for Figure 3a \hfil }
\begin{center} 
\begin{tabular}{lcccccc}
\noalign{\smallskip}
\hline\hline
\noalign{\smallskip}
 $\Sigma$  &  f(E)~~~$\sigma$  &  f(S0)~~~$\sigma$ &  f(Sp)~~~$\sigma$ & \ \cr
\noalign{\smallskip}
\noalign{\hrule}
\noalign{\smallskip}
  0.35  & 0.087  0.137  &  0.447  0.215  &  0.466  0.221  \cr
  0.55  & 0.043  0.051  &  0.537  0.130  &  0.420  0.130  \cr
  0.75  & 0.121  0.050  &  0.458  0.076  &  0.421  0.076  \cr
  0.95  & 0.197  0.044  &  0.414  0.054  &  0.389  0.054  \cr
  1.15  & 0.250  0.037  &  0.463  0.043  &  0.287  0.040  \cr
  1.35  & 0.198  0.026  &  0.527  0.033  &  0.276  0.030  \cr
  1.55  & 0.243  0.026  &  0.509  0.031  &  0.248  0.027  \cr
  1.75  & 0.302  0.030  &  0.430  0.033  &  0.268  0.029  \cr
  1.95  & 0.394  0.039  &  0.438  0.040  &  0.169  0.030  \cr
  2.15  & 0.367  0.057  &  0.550  0.059  &  0.084  0.033  \cr
  2.35  & 0.409  0.105  &  0.546  0.106  &  0.045  0.045  \cr
\end{tabular}
\end{center}

\centerline{\sc \hfil Data for Figure 3b \hfil }
\begin{center} 
\begin{tabular}{lcccccc}
\noalign{\smallskip}
\hline\hline
\noalign{\smallskip}
 $\Sigma$  &  f(E)~~~$\sigma$  &  f(S0)~~~$\sigma$ &  f(Sp)~~~$\sigma$ & \ \cr
\noalign{\smallskip}
\noalign{\hrule}
\noalign{\smallskip}
  0.55  & 0.000  0.000  &  0.524  0.298  &  0.476  0.298  \cr
  0.75  & 0.096  0.100  &  0.375  0.170  &  0.529  0.177  \cr
  0.95  & 0.099  0.103  &  0.517  0.171  &  0.384  0.169  \cr
  1.15  & 0.308  0.132  &  0.542  0.143  &  0.149  0.105  \cr
  1.35  & 0.174  0.051  &  0.577  0.068  &  0.249  0.060  \cr
  1.55  & 0.202  0.044  &  0.546  0.055  &  0.252  0.049  \cr
  1.75  & 0.331  0.048  &  0.492  0.051  &  0.177  0.039  \cr
  1.95  & 0.432  0.070  &  0.433  0.070  &  0.135  0.049  \cr
  2.15  & 0.263  0.101  &  0.685  0.107  &  0.052  0.052  \cr
  2.35  & 0.364  0.146  &  0.547  0.151  &  0.089  0.088  \cr
\end{tabular}
\end{center}

\centerline{\sc \hfil Data for Figure 3c \hfil }
\begin{center} 
\begin{tabular}{lcccccc}
\noalign{\smallskip}
\hline\hline
\noalign{\smallskip}
 $\Sigma$  &  f(E)~~~$\sigma$  &  f(S0)~~~$\sigma$ &  f(Sp)~~~$\sigma$ & \ \cr
\noalign{\smallskip}
\noalign{\hrule}
\noalign{\smallskip}
  0.95  & 0.168  0.163  &  0.344  0.205  &  0.489  0.219  \cr
  1.15  & 0.158  0.062  &  0.509  0.085  &  0.333  0.081  \cr
  1.35  & 0.216  0.070  &  0.492  0.085  &  0.292  0.078  \cr
  1.55  & 0.159  0.060  &  0.476  0.082  &  0.365  0.080  \cr
  1.75  & 0.335  0.112  &  0.334  0.113  &  0.331  0.113  \cr
  1.95  & 0.480  0.105  &  0.261  0.092  &  0.259  0.092  \cr
  2.15  & 0.501  0.251  &  0.499  0.251  &  0.000  0.000  \cr
\end{tabular}
\end{center}

\newpage
\centerline{\sc \hfil Data for Figure 4 \hfil }
\begin{center} 
\begin{tabular}{lcccccc}
\noalign{\smallskip}
\hline\hline
\noalign{\smallskip}
 $\Sigma$  &  f(E)~~~$\sigma$  &  f(S0)~~~$\sigma$ &  f(Sp)~~~$\sigma$ & \ \cr
\noalign{\smallskip}
\noalign{\hrule}
\noalign{\smallskip}
  1.35  & 0.437  0.159  &  0.137  0.090  &  0.426  0.186  \cr
  1.55  & 0.355  0.100  &  0.238  0.087  &  0.407  0.126  \cr
  1.75  & 0.334  0.054  &  0.138  0.039  &  0.528  0.061  \cr
  1.95  & 0.318  0.036  &  0.184  0.030  &  0.498  0.041  \cr
  2.15  & 0.370  0.033  &  0.182  0.026  &  0.448  0.036  \cr
  2.35  & 0.478  0.041  &  0.216  0.033  &  0.305  0.041  \cr
  2.55  & 0.485  0.055  &  0.184  0.041  &  0.331  0.055  \cr
  2.75  & 0.770  0.084  &  0.230  0.084  &  0.000  0.000  \cr
\end{tabular}
\end{center}

\centerline{\sc \hfil Data for Figure 6 \hfil }
\begin{center} 
\begin{tabular}{lcccccc}
\noalign{\smallskip}
\hline\hline
\noalign{\smallskip}
 $\Sigma$  &  f(E)~~~$\sigma$  &  f(S0)~~~$\sigma$ &  f(Sp)~~~$\sigma$ & \ \cr
\noalign{\smallskip}
\noalign{\hrule}
\noalign{\smallskip}
  1.35  & 0.586  0.458  &  0.105  0.156  &  0.309  0.514  \cr
  1.55  & 0.426  0.233  &  0.184  0.152  &  0.390  0.281  \cr
  1.75  & 0.320  0.074  &  0.145  0.054  &  0.534  0.085  \cr
  1.95  & 0.350  0.051  &  0.198  0.043  &  0.452  0.058  \cr
  2.15  & 0.306  0.046  &  0.202  0.040  &  0.491  0.052  \cr
  2.35  & 0.450  0.068  &  0.190  0.053  &  0.359  0.069  \cr
  2.55  & 0.292  0.081  &  0.235  0.075  &  0.473  0.092  \cr
\end{tabular}
\end{center}

\centerline{\sc \hfil Data for Figure 8 \hfil }
\begin{center} 
\begin{tabular}{lcccccc}
\noalign{\smallskip}
\hline\hline
\noalign{\smallskip}
 $\Sigma$  &  f(E)~~~$\sigma$  &  f(S0)~~~$\sigma$ &  f(Sp)~~~$\sigma$ & \ \cr
\noalign{\smallskip}
\noalign{\hrule}
\noalign{\smallskip}
  1.35  & 0.260  0.166  &  0.176  0.149  &  0.564  0.217  \cr
  1.55  & 0.317  0.120  &  0.264  0.119  &  0.419  0.158  \cr
  1.75  & 0.388  0.102  &  0.134  0.068  &  0.478  0.113  \cr
  1.95  & 0.315  0.060  &  0.157  0.048  &  0.528  0.069  \cr
  2.15  & 0.447  0.055  &  0.187  0.041  &  0.366  0.059  \cr
  2.35  & 0.520  0.055  &  0.247  0.046  &  0.233  0.052  \cr
  2.55  & 0.603  0.068  &  0.152  0.047  &  0.245  0.064  \cr
  2.75  & 0.751  0.090  &  0.249  0.090  &  0.000  0.000  \cr
\end{tabular}
\end{center}

\smallskip

\newpage
\vskip0.5truein
\centerline{\hbox{\psfig{figure=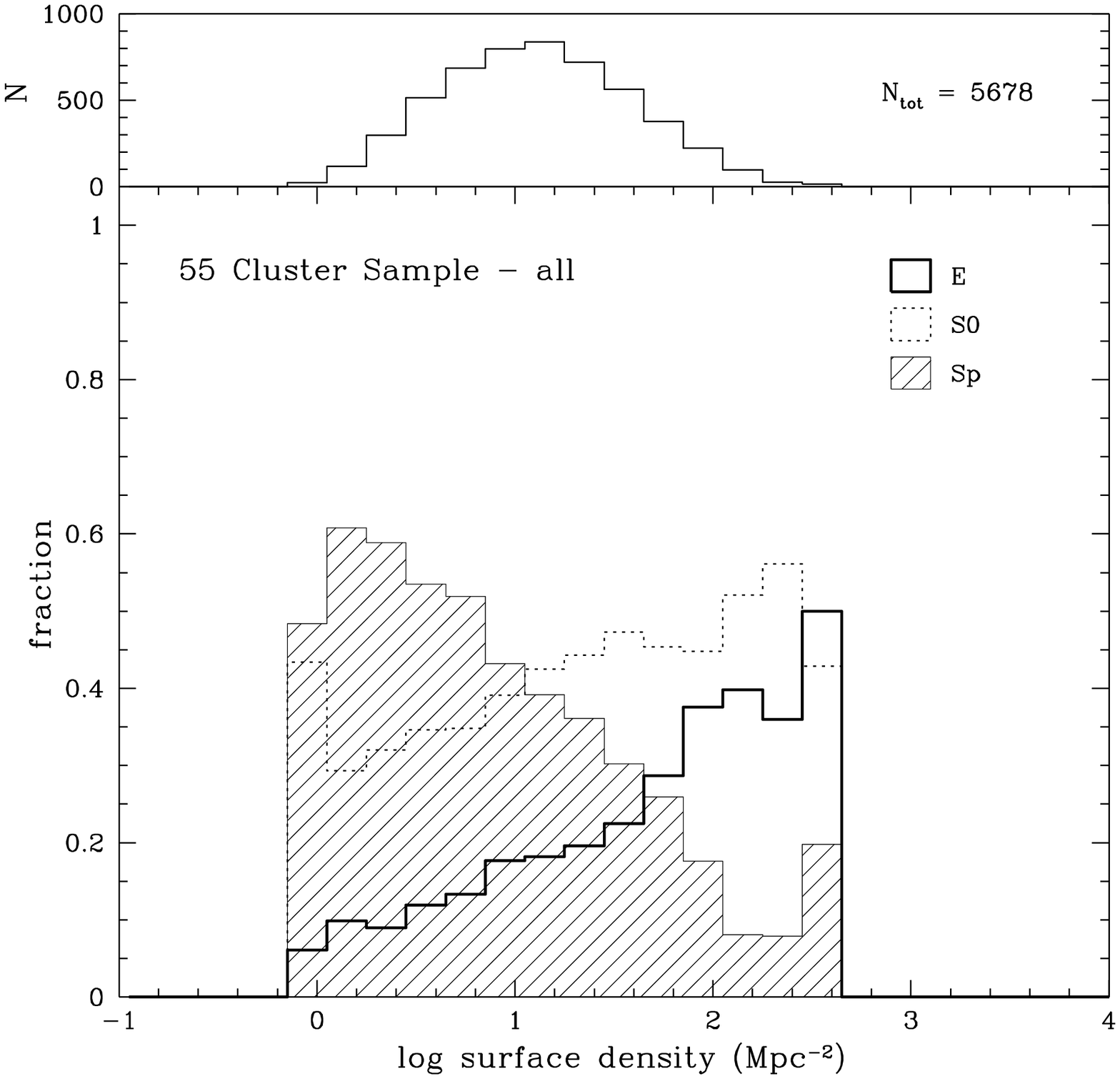,width=6.0in}}}
{{\bf Figure 1.} The \MD (morphology-density) relation for the D80 
55 cluster sample, reanalyzed, as discussed in the text. 
The histogram at the top shows the numbers of galaxies in each bin 
of surface density, for the total number in the upper right corner.
}

\newpage
\vskip0.5truein
\centerline{\hbox{\psfig{figure=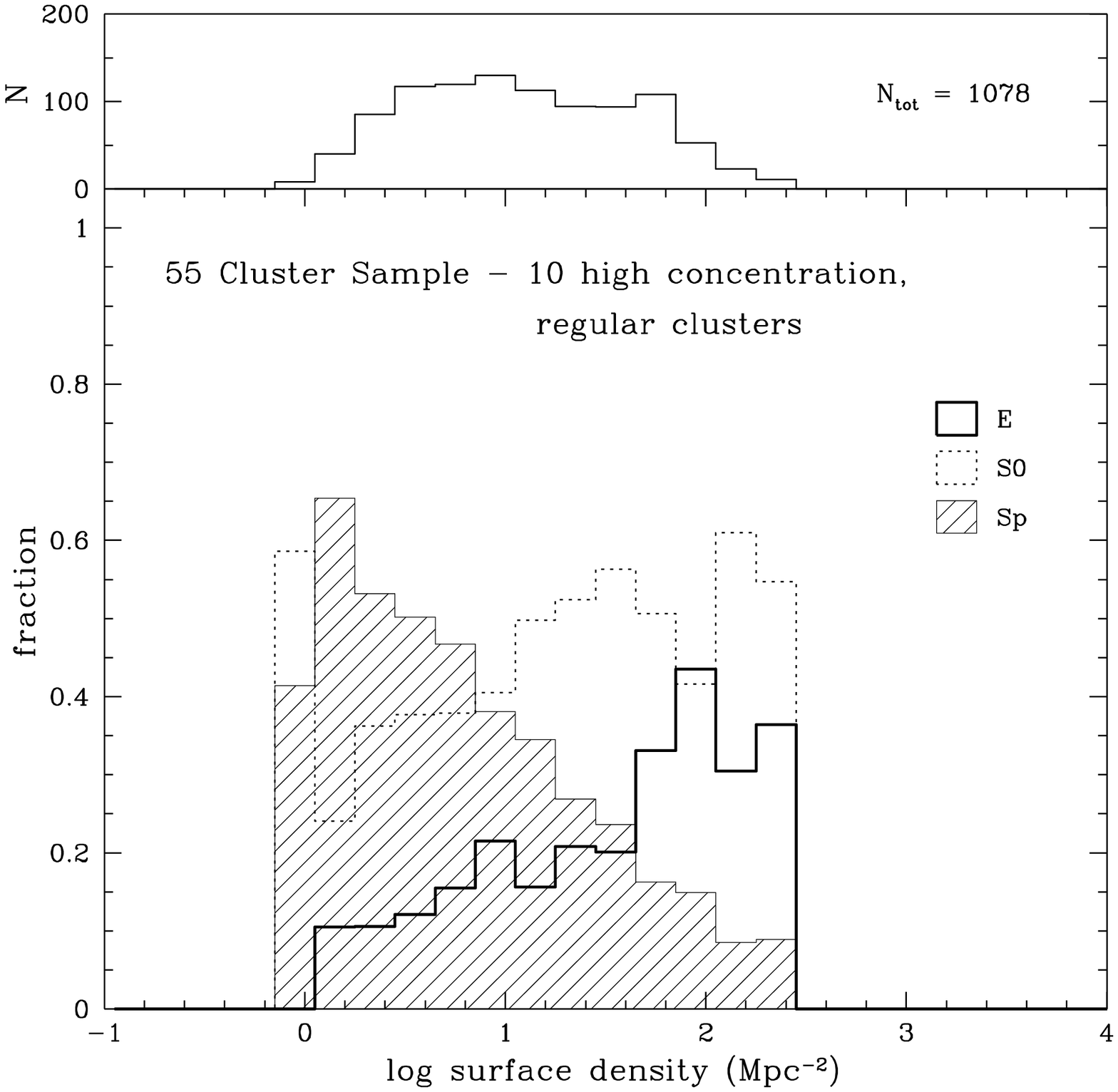,width=3.0in}}}
\centerline{\hbox{\psfig{figure=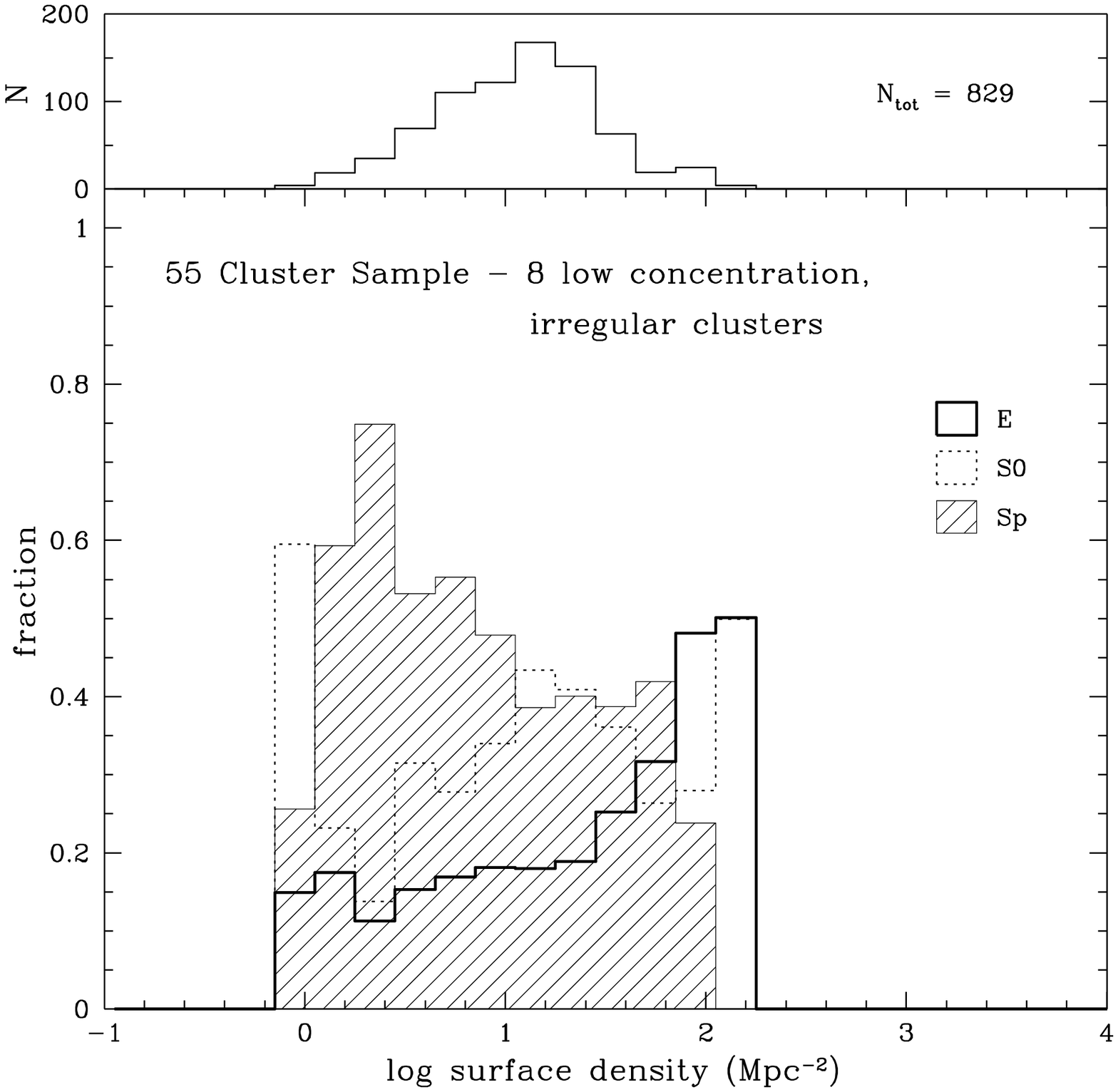,width=3.0in}}}
{{\bf Figure 2.} The \MD relation for 10 centrally concentrated, regular
clusters of the D80 cluster sample (above), A151, A539, A957,
A1656, A1913, A2040, A2063, DC0247-31, DC0428-53, DC1842-63, and
(below), 8 low-concentration clusters A76, A119, A168, A978, A979,
A1644, A2151, and DC0003-50.
}

\newpage
\vskip0.5truein
\centerline{\hbox{\psfig{figure=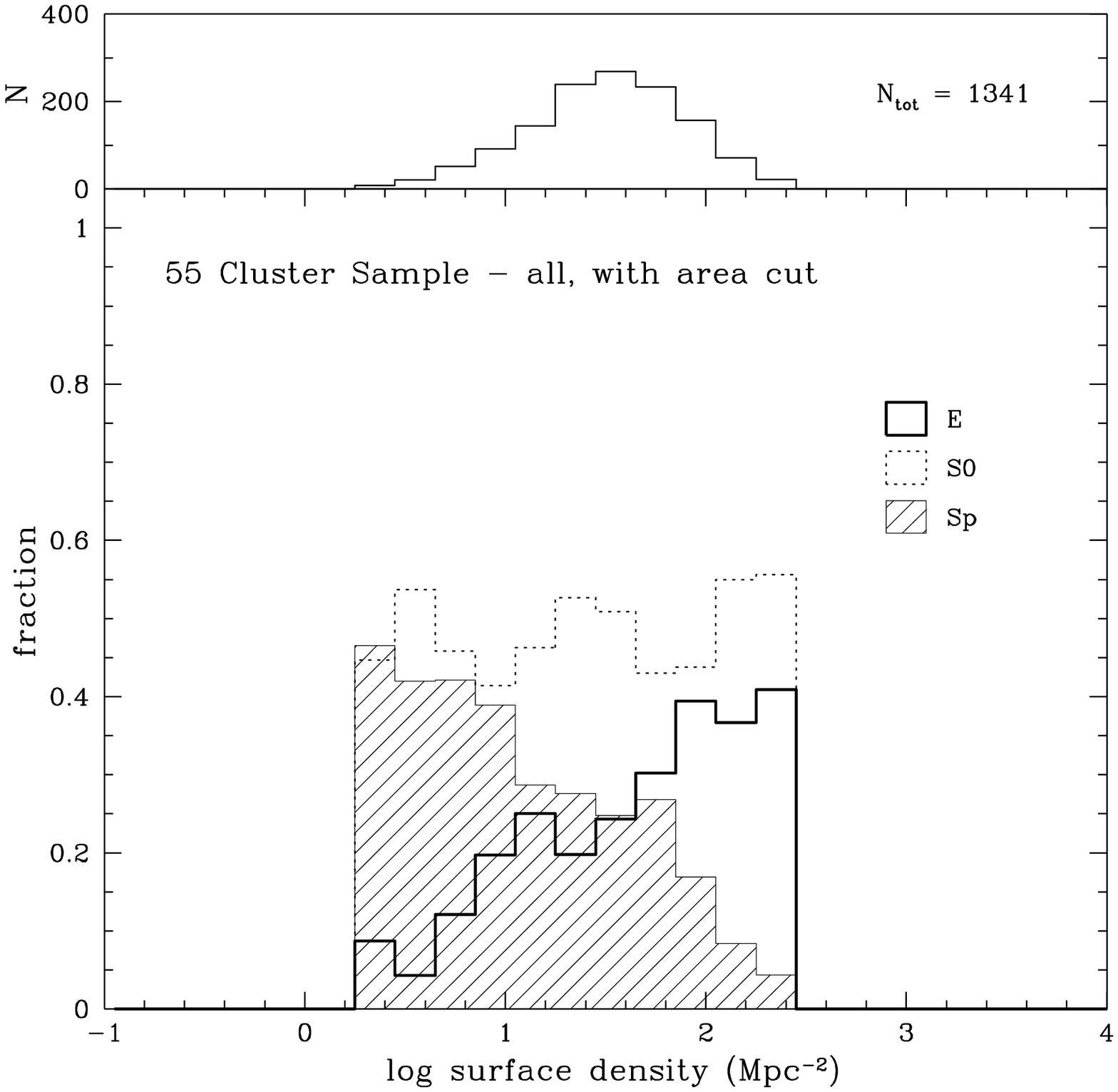,width=2.0in}}}
\centerline{\hbox{\psfig{figure=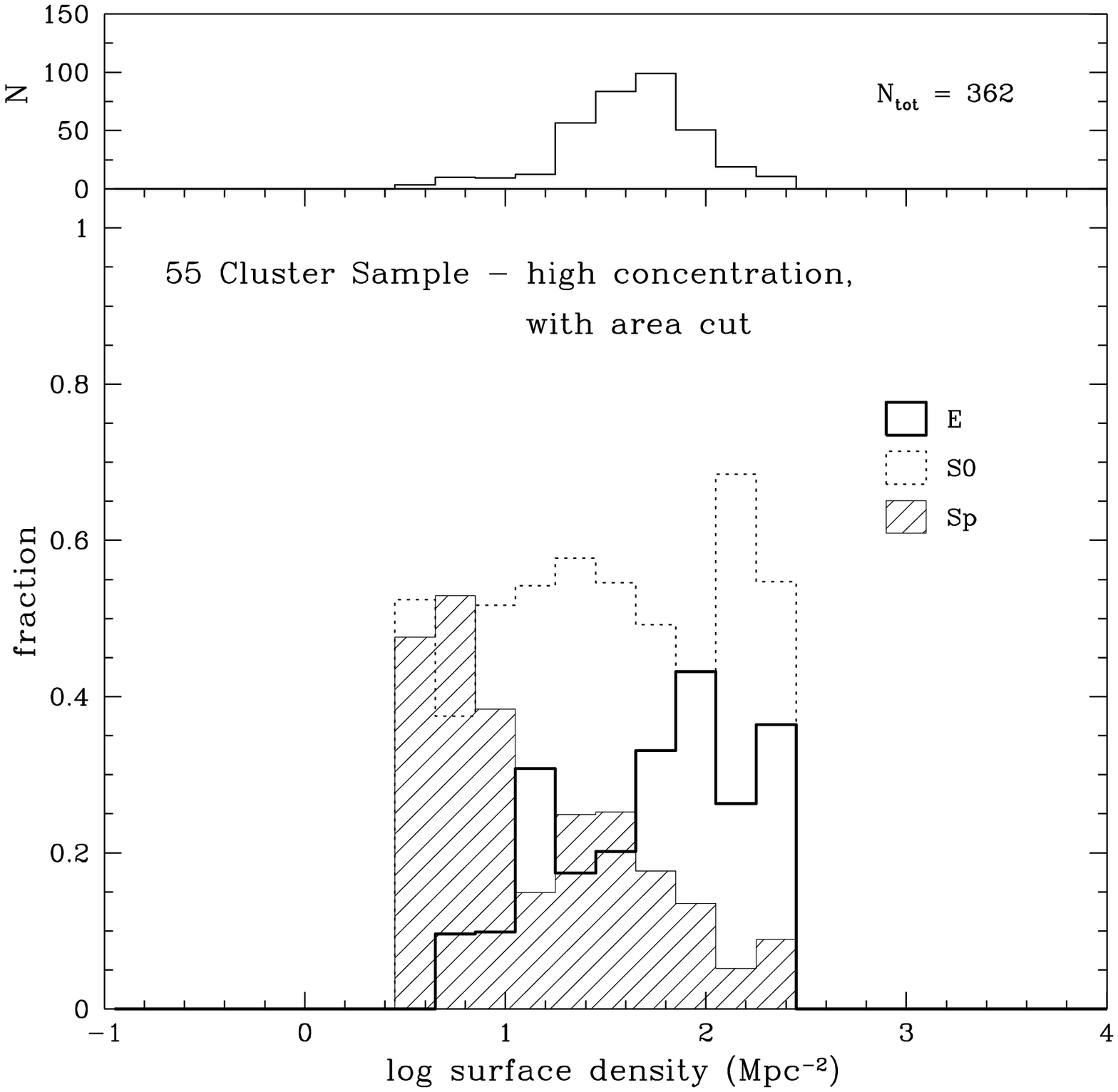,width=2.0in}}}
\centerline{\hbox{\psfig{figure=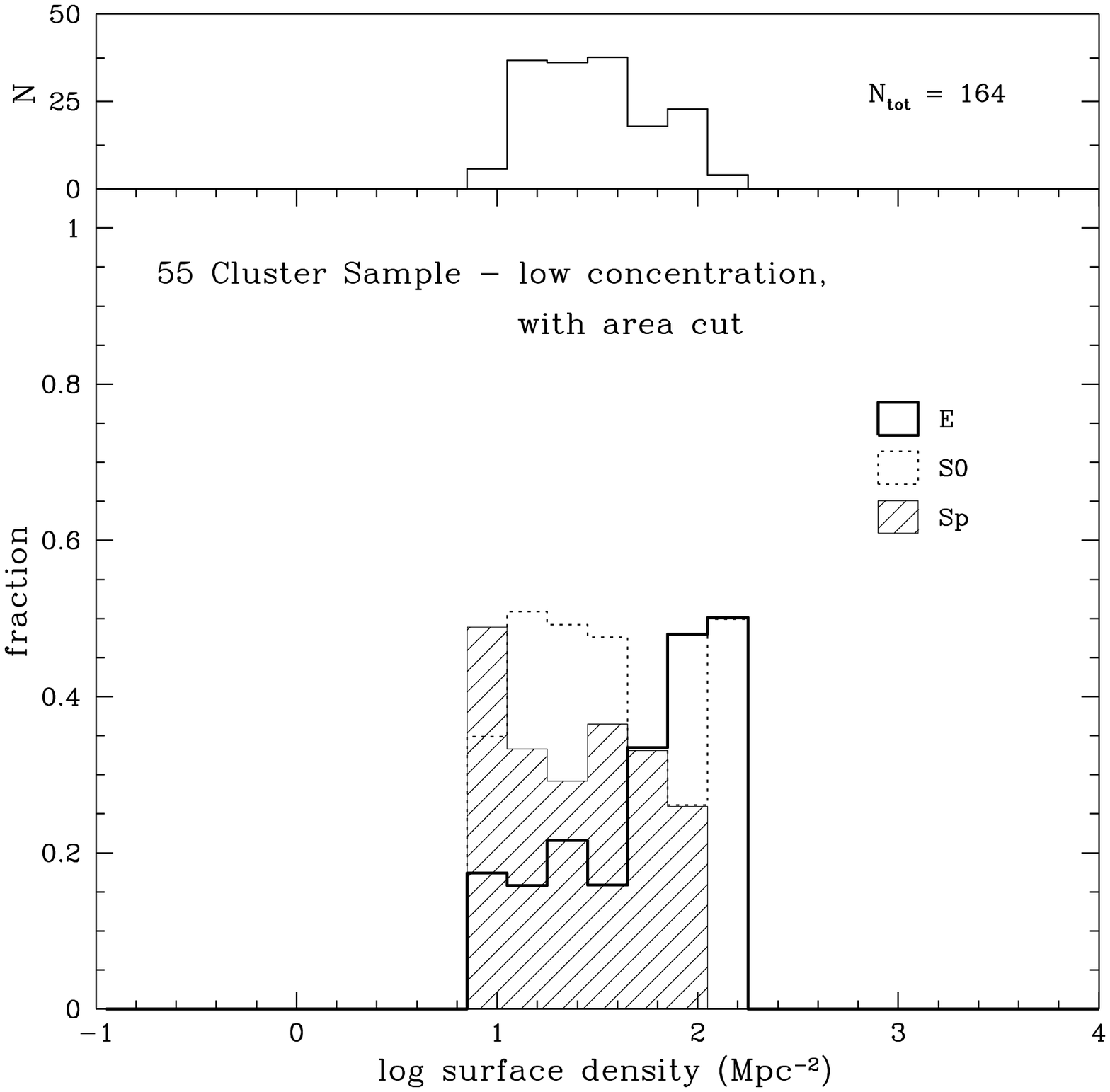,width=2.0in}}}
{{\bf Figure 3.} 
The \MD relation for all 55 clusters of the D80 cluster sample, 
but limited to a smaller area of 1.2 Mpc, comparable to that covered
by the more distant HST sample.  From top to bottom:  all clusters;
centrally-concentrated clusters, low-concentration clusters.
}

\newpage
\vskip0.5truein
\centerline{\hbox{\psfig{figure=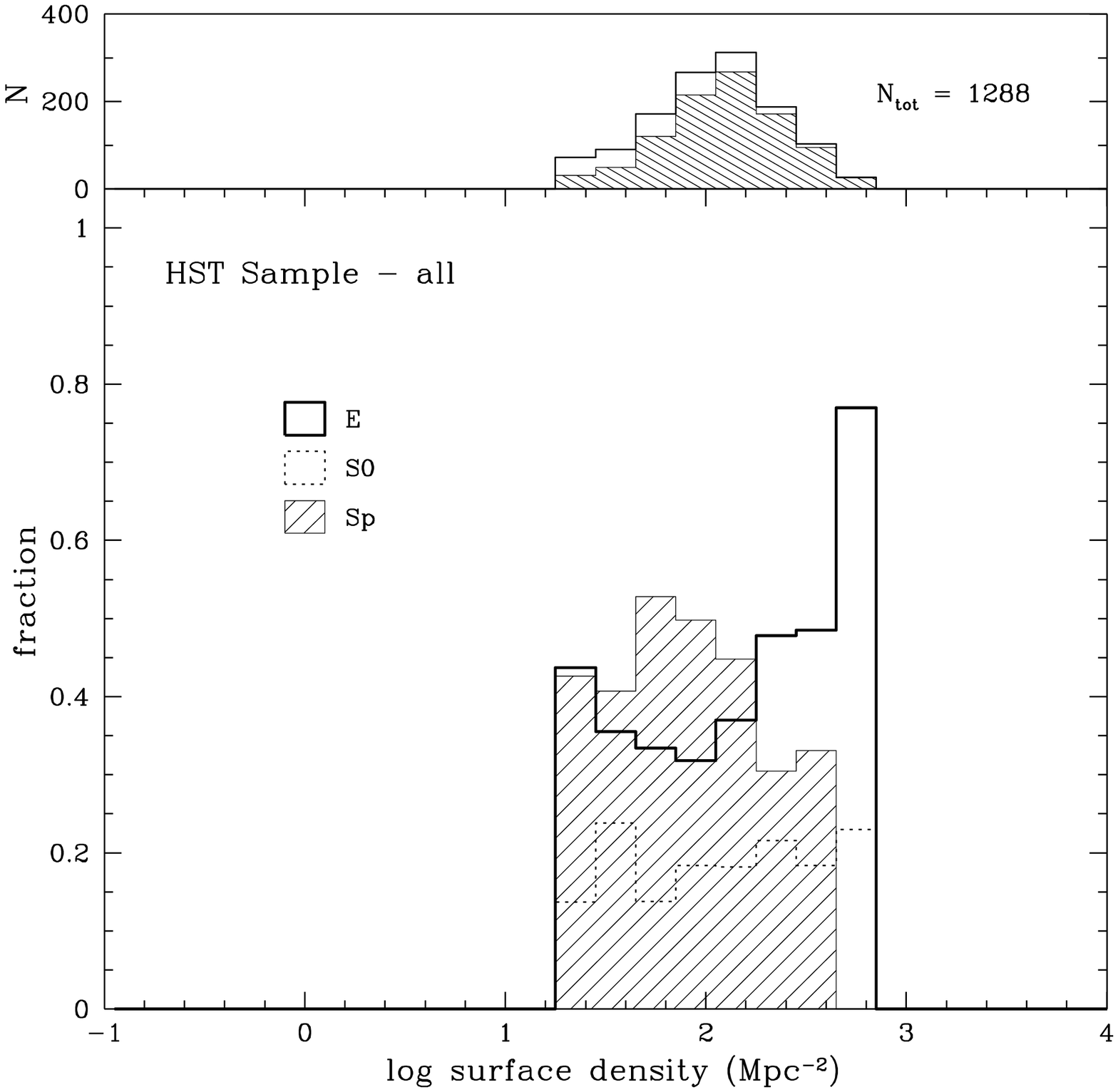,width=6.0in}}}
{{\bf Figure 4.} 
The \MD relation for 10 clusters at intermediate redshift, 
$0.36 < z < 0.57$.}  The histogram in the top box shows the
number of galaxies in each bin, with the shaded portion indicating
the number after correcting for field contamination.

\newpage
\vskip0.5truein
\centerline{\hbox{\psfig{figure=fig5_md.ps,width=6.0in}}}
{{\bf Figure 5.} 
The distribution of ellipticities for E and S0 galaxies in
the $z \sim 0.5$ sample compared to that for the Coma cluster (Andreon
et al.\ 1996) and for 11 clusters $0.035 < z < 0.044$ of the Dressler
(1980) low redshift sample.  The similarity of these distributions
suggests that a fair comparison of the fraction of E and S0 galaxies
can be made between $z \sim 0$ and $z \sim 0.5$.}

\newpage
\vskip0.5truein
\centerline{\hbox{\psfig{figure=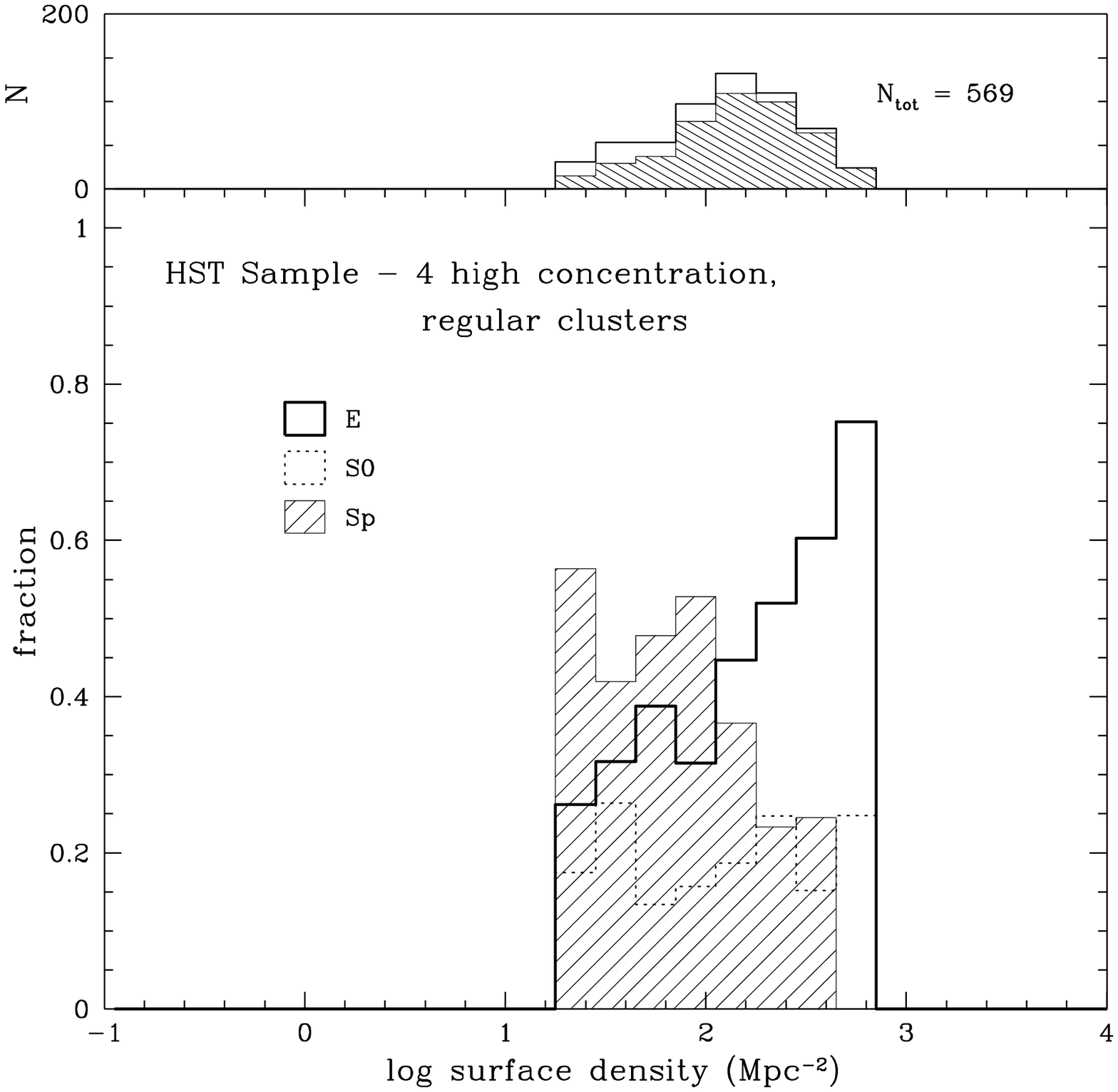,width=6.0in}}}
{{\bf Figure 6.} 
The \MD relation for 4 centrally concentrated, 
regular clusters at intermediate redshift, 3C295, Cl0016+16, Cl0024+16, 
and Cl0054$-$27.}

\newpage
\vskip0.5truein
\centerline{\hbox{\psfig{figure=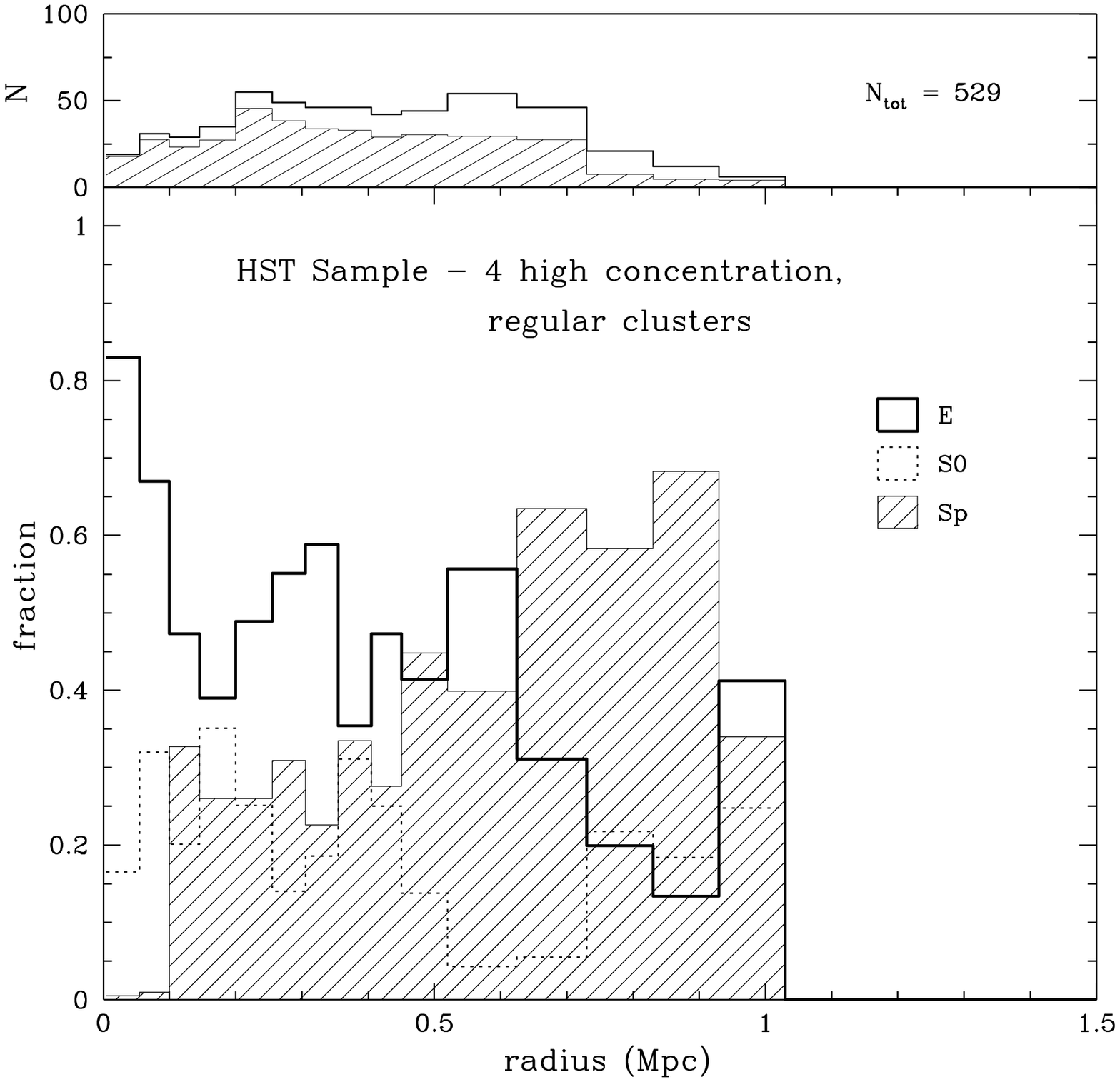,width=6.0in}}}
{{\bf Figure 7.} 
The T-R (morphology-clustocentric radius) 
relation for 4 high-concentration, regular clusters at intermediate 
redshift, 3C295, Cl0016+16, Cl0024+16, and Cl0054$-$27.}

\newpage
\vskip0.5truein
\centerline{\hbox{\psfig{figure=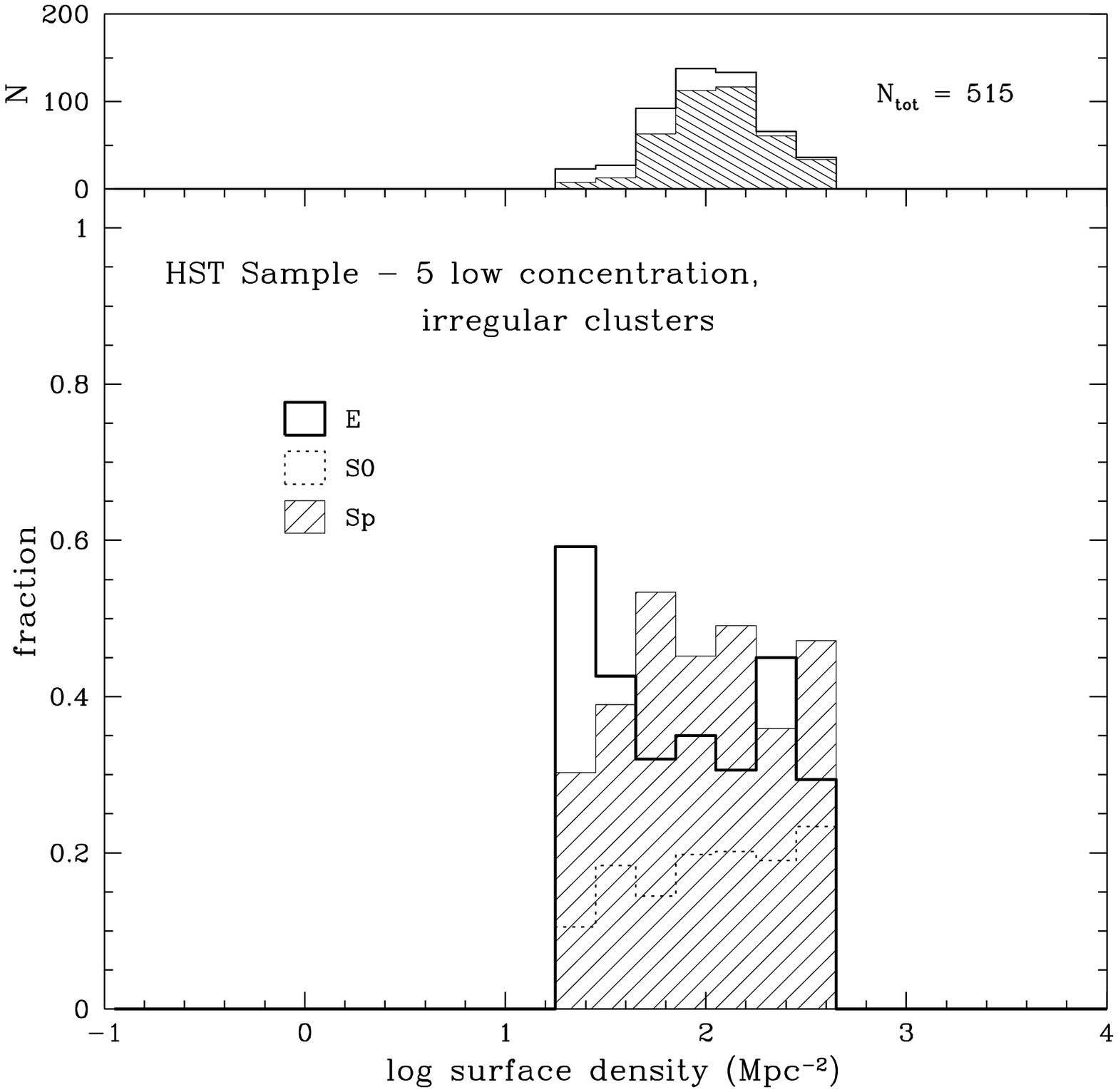,width=6.0in}}}
{{\bf Figure 8.} The \MD relation for 5 low-concentration, 
irregular clusters at intermediate redshift, Cl0303+17, Cl0412$-$65, 
Cl0939+47, Cl1447+23, and Cl1601+42.
}

\newpage
\vskip0.5truein
\centerline{\hbox{\psfig{figure=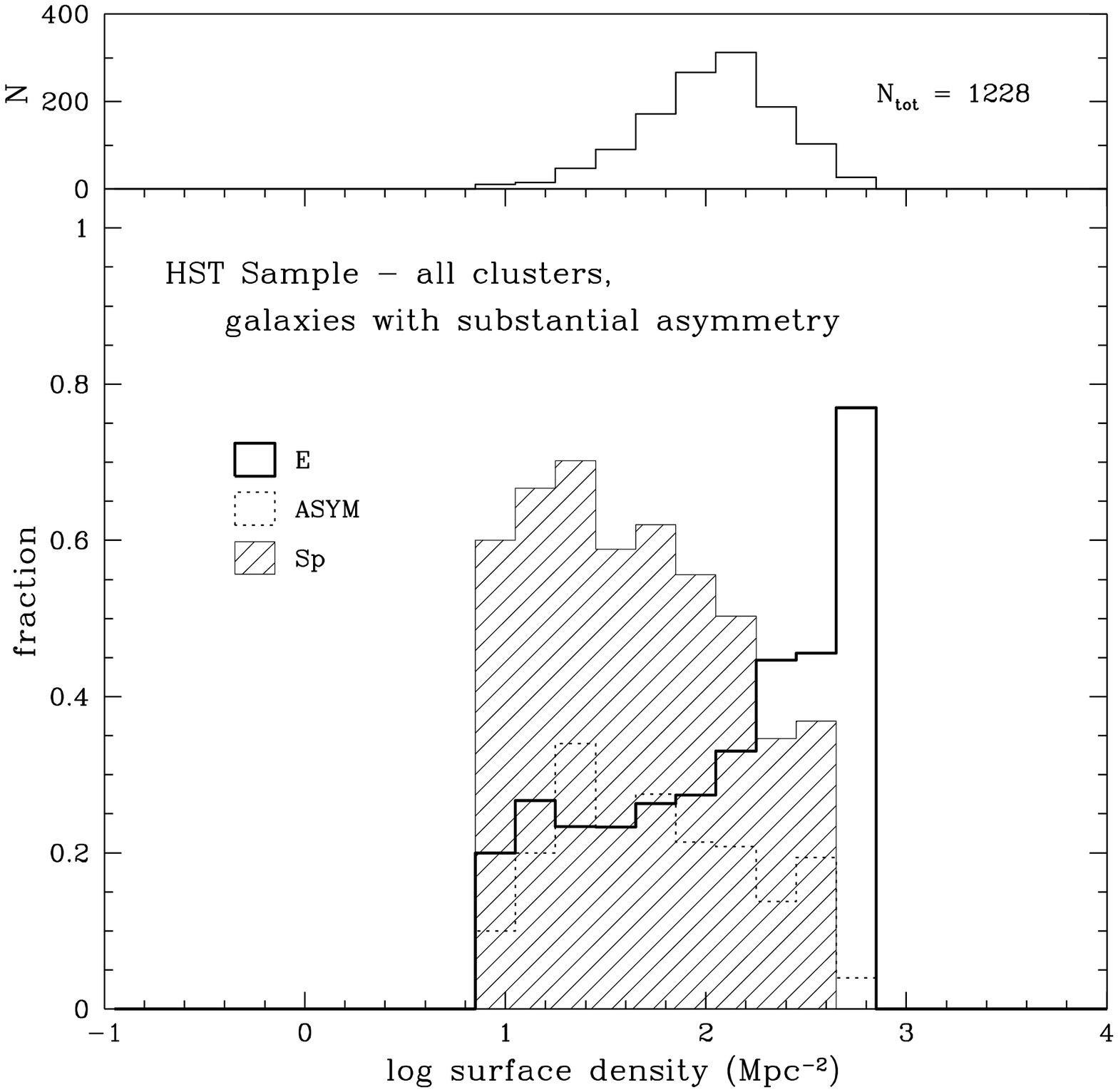,width=6.0in}}}
{{\bf Figure 9.} The \MD relation for 10 clusters at intermediate redshift, 
including those galaxies with substantial asymmetry or disturbance, irrespective of morphological type.  These galaxies follow the spiral or
S0 trend rather than the elliptical trend.
}

\newpage
\vskip0.5truein
\centerline{\hbox{\psfig{figure=fig10_md.ps,width=6.0in}}}
{{\bf Figure 10.} The S0/E fraction for clusters in the sample as a function of
redshift.  The open circles are the outer fields in A370 and Cl0939+47,
which are not used in the least-squares fit, shown with its 1-$\sigma$
errors as the solid and dotted lines.  The extrapolation of this linear
relation to zero redshift approximately matches the value S0/E $\sim$ 2,
shown by the solid box, found for 11 clusters $0.035 < z < 0.044$ of the D80 cluster sample.
}

\newpage
\centerline{\hbox{\psfig{file=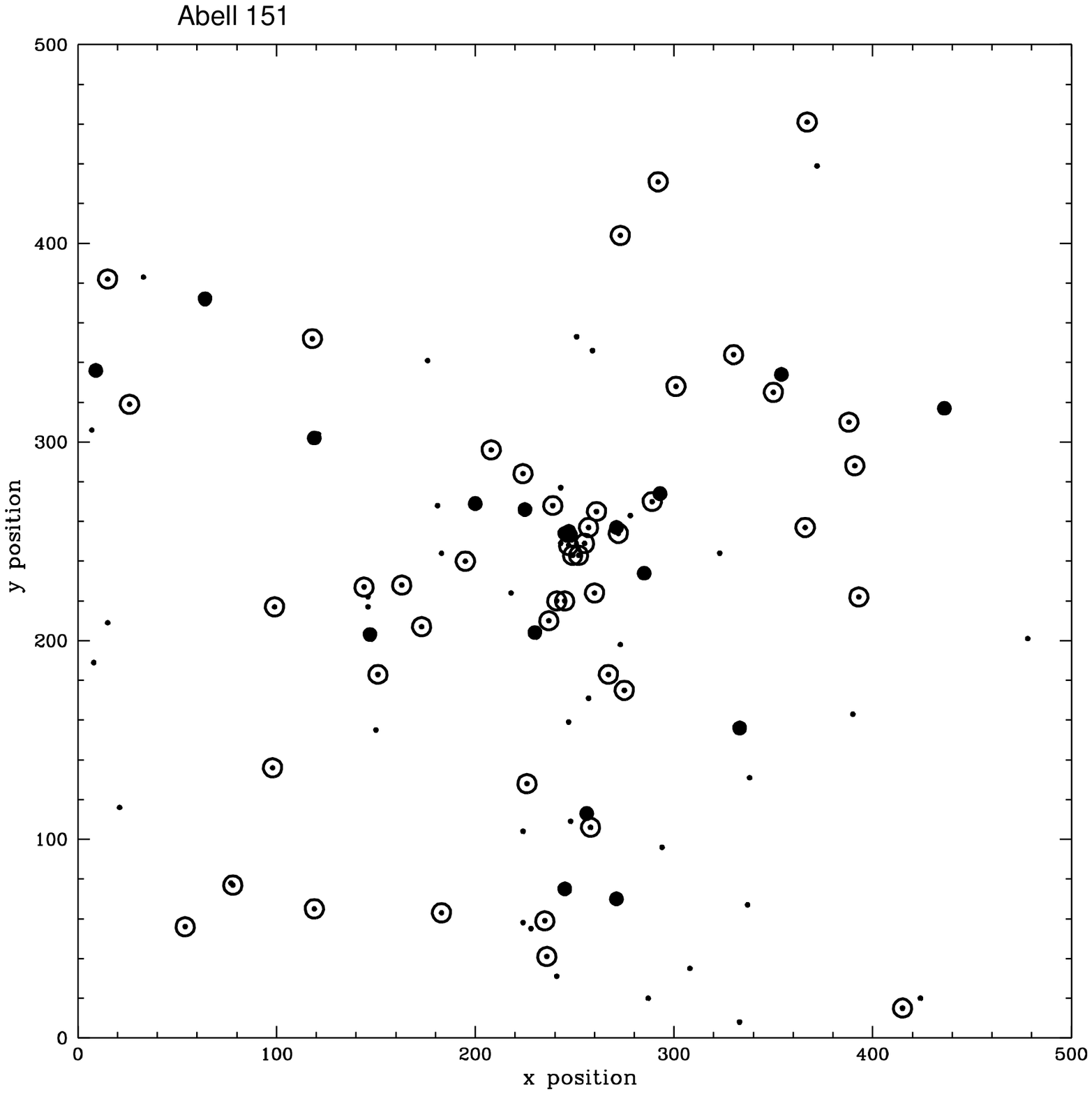,width=3.0in}\psfig{file=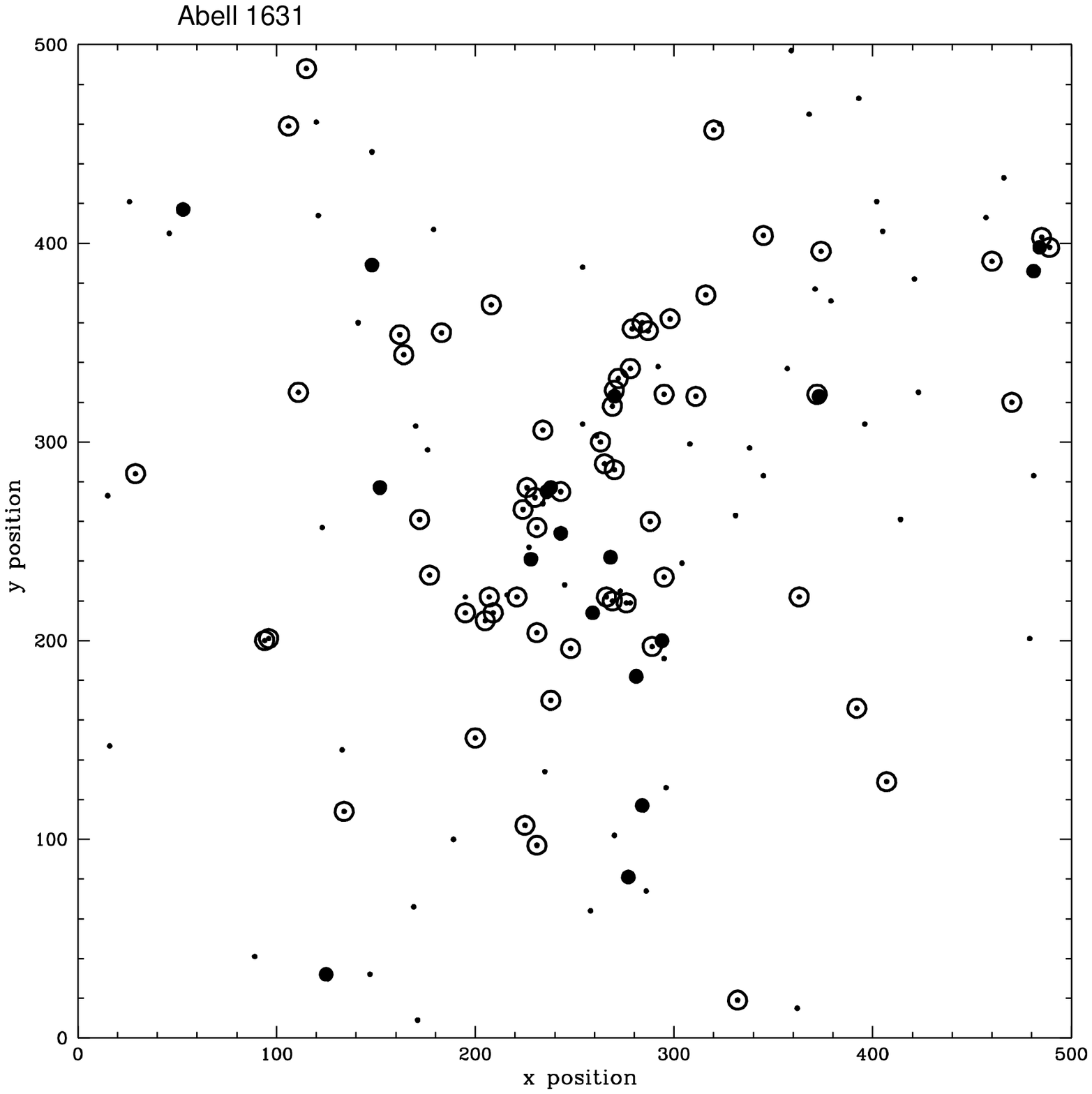,width=3.0in}}}
\centerline{\hbox{\psfig{file=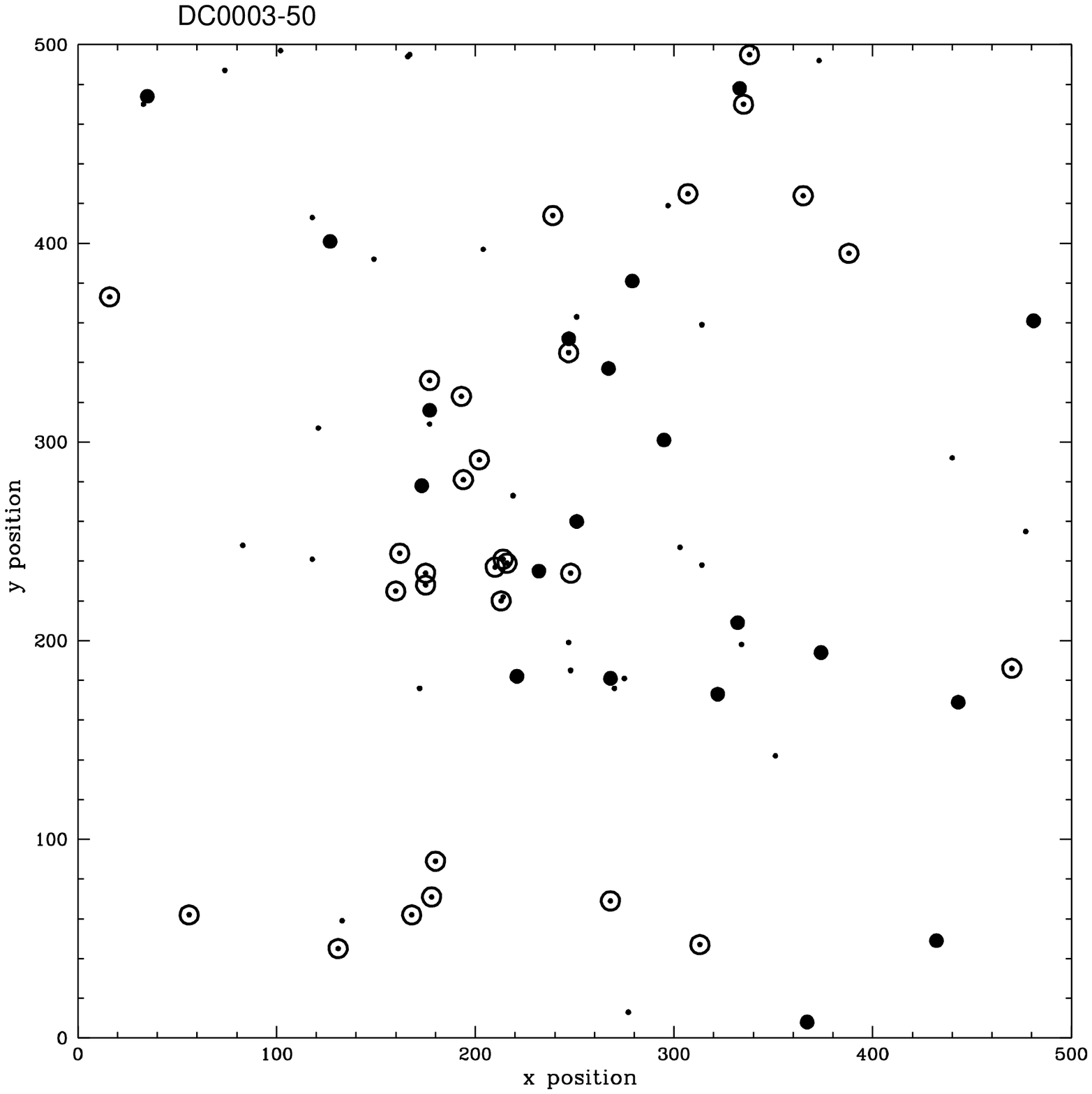,width=3.0in}\psfig{file=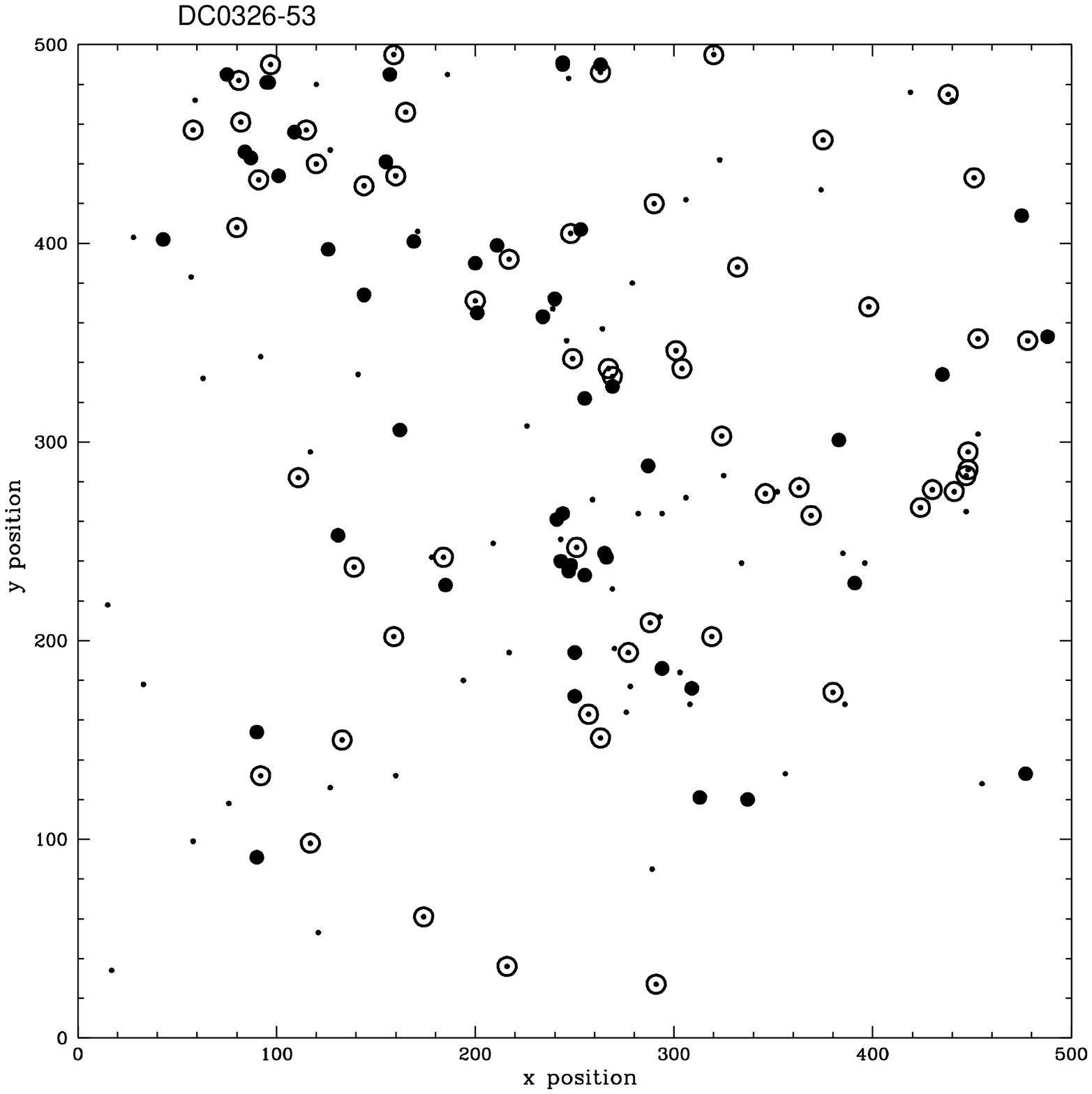,width=3.0in}}}

{{\bf Figure 11.} Maps showing the positions of galaxies in four clusters 
of the D80 55 cluster sample, A151, A1631, DC0003-50, and DC0326-53.
Large solid dots are elliptical galaxies, dotted-circles are S0's,
and small dots are spirals.  The enhanced clustering of E and S0
galaxies compared to spirals is evident.
}

\newpage
\vskip0.5truein
\centerline{\hbox{\psfig{figure=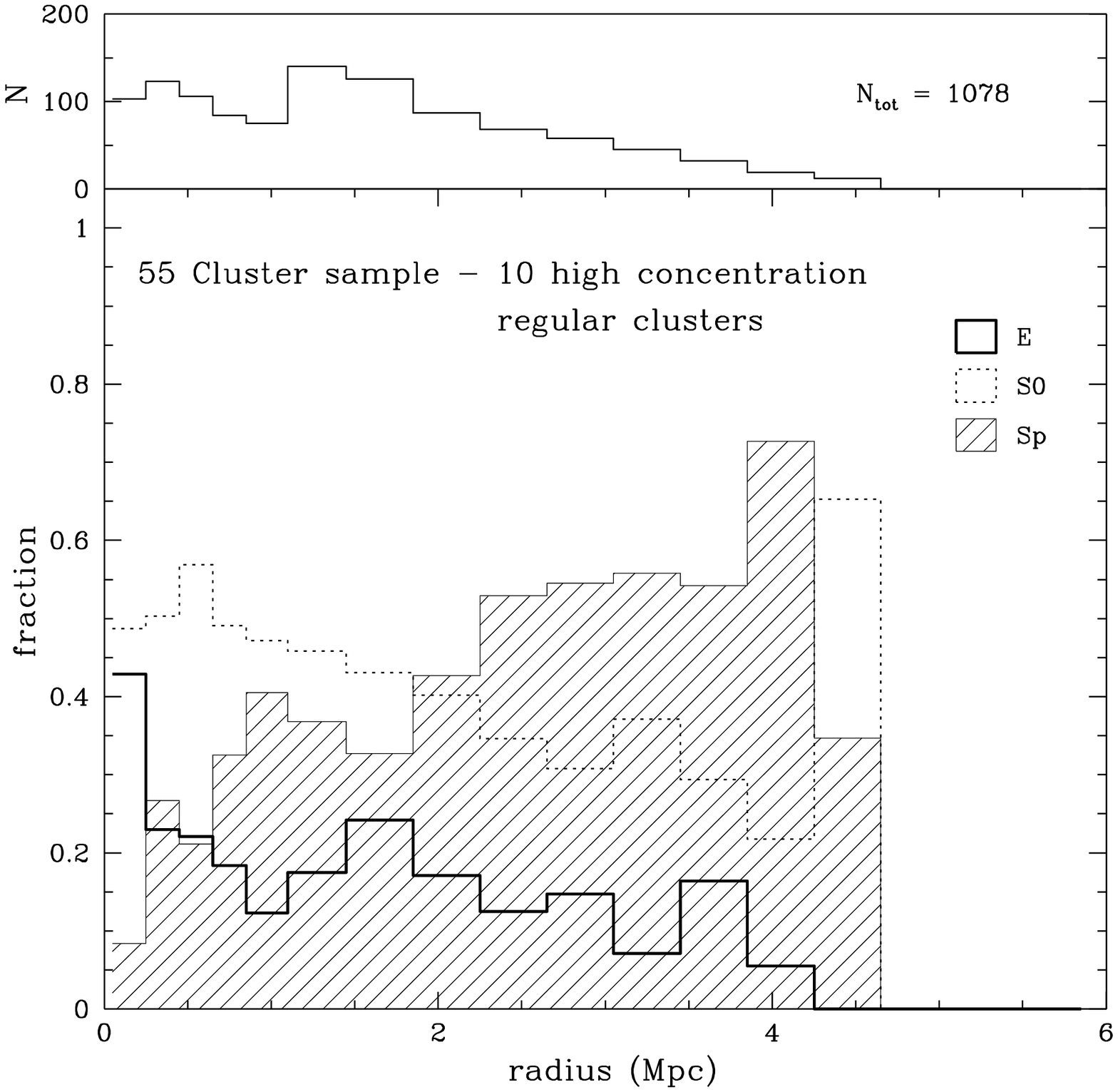,width=3.0in}}}
\centerline{\hbox{\psfig{figure=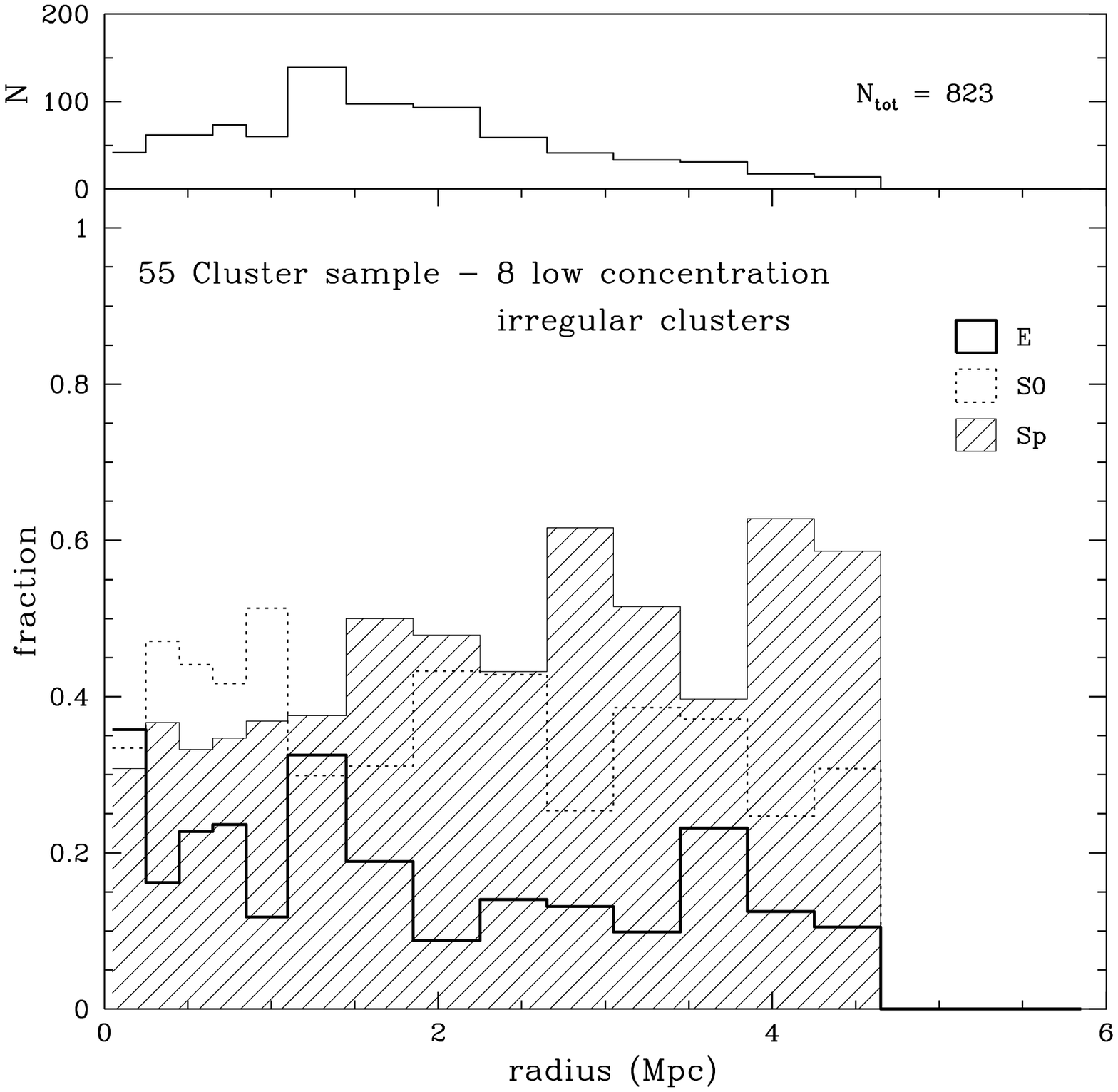,width=3.0in}}}
{{\bf Figure 12.} The T-R relation for the same clusters of Fig.~3. 
As expected, population gradients are strong for centrally concentrated 
clusters (above), and weak or absent for low-concentration irregular 
clusters.
}


\newpage
\begin{thebibliography}{}
\itemsep=0in

\bibitem[Andreon et al.\ 1996]{a96}
Andreon, S., Davoust, E., Michard, R., Nieto, J.-L. \& Poulain, P. 1996
A\&AS, 116, 429.

\bibitem[Beers \& Tonry 1986]{bt86}
Beers, T.C. \& Tonry, J.L. 1986, ApJ, 300, 557.

\bibitem[Bertin \& Arnoults 1996]{ba96} 
Bertin E., \& Arnoults, S. 1996, A\&AS, 117, 393.

\bibitem[Bhavsar 1981]{bh81} Bhavsar, S.P., 1981, ApJL, 246, L5.

\bibitem[Bower, Lucey \& Ellis 1992]{ble92}
Bower, R.G., Lucey, J.R.\ \& Ellis, R.S., 1991, MNRAS, 254, 601.

\bibitem[Butcher \& Oemler 1978]{bo78}
Butcher, H., \& Oemler, A. Jr. 1978, ApJ, 226, 559. 

\bibitem[de Sousa et al.\ 1982]{ds82} de Sousa, R.E., Capelato, H.V., 
Arakaki, L, \& Loguillo, C., 1982, ApJ, 263, 557.

\bibitem[Dressler 1980]{d80} Dressler, A. 1980a, ApJ, 236, 351. (D80)

\bibitem[Dressler 1980b]{d80b} Dressler, A. 1980b, ApJS, 424, 565. (DCAT80)

\bibitem[Ellis et al.\ 1996]{e96}
Ellis, R.S., Smail, I., Dressler, A., Couch. W.J., Oemler, A. Jr.,
Butcher, H., \& Sharples, R.M. 1996, ApJ, in press.

\bibitem[Evrard, Silk, \& Szalay 1990]{ess90}
Evrard, A., Silk, J., \& Szalay, A.S. 1990, ApJ, 365, 13.

\bibitem[Faber \& Gallagher 1976]{fg76} Faber, S.M, \& Gallagher, J.S.,
1976, ApJ, 204, 668.

\bibitem[Geller \& Beers 1982]{gb82}
Geller, M.J, and Beers, T.C. 1982, PASP, 94, 421.

\bibitem[Giovanelli et al.\ 1986]{ghc} Giovanelli, R., Haynes, M.P., \&
Chincarini, G.L. 1986, ApJ, 300, 77.

\bibitem[Griffiths et al.\ 1994]{grif94}
Griffiths, R.E., Casertano, S., Ratnatunga, K.U., Neuschaefer, L.W.,
Ellis, R.S., Gilmore, G.F., Glazebrook, K. Santiago, B., Huchra, J.P.,
et al. 1994, ApJ, 435, L19.

\bibitem[Gunn \& Gott 1972]{gg72}
Gunn, J.E., \& Gott, J.R. 1972, Ap.J., 176, 1.

\bibitem[Holtzman et al.\ 1995]{h95}	
Holtzman, J.A., Burrows, C.J., Casertano, S., Hester, J.J.,
Trauger, J.T., Watson, A.M.\ \& Worthey, G. 1995, PASP, 107, 1065.

\bibitem[Hubble \& Humason 1931]{hh31} Hubble, E., \& Humason, M.L. 1931,
ApJ, 74, 43.

\bibitem[Melnick \& Sargent 1977]{ms77} Melnick, J., \& Sargent, W.L.W.,
1977, ApJ, 215, 401. 

\bibitem[Moore et al. 1996]{m96}
Moore, B., Katz, N., Lake, G., Dressler, A., and Oemler, A. Jr. 1996,
Nature, 379, 613.

\bibitem[Oemler 1974]{o74} Oemler, A. Jr., 1974, ApJ 194, 1.

\bibitem[Oemler, Dressler, and Butcher]{odb97}
Oemler, A. Jr., Dressler, A., and Butcher, H. 1997, ApJ, 474, 561.

\bibitem[Postman \& Geller 1984]{pg84} Postman, M., \& Geller, M.J., 1984,
ApJ, 281, 95.

\bibitem[Salvador-Sole et al. 1989]{s89}
Salvador-Sole, E. Sanroma, M., and Rdz.Jordana, J.J. 1989, ApJ, 337, 636.

\bibitem[Sandage, Binggeli, \& Tammann 1985]{sbt85} 
Sandage, A., Binggeli, B., and Tammann, G.A. 1985, AJ, 90, 395.

\bibitem[Sandage, Freeman, \& Stokes 1970]{sfs70}
Sandage, A., Freeman, K.C., \& Stokes, N.R. 1970, ApJ, 160, 831.

\bibitem[Sanroma \& Salvador-Sole 1990]{ss90} Sanroma, M., \& Salvador-Sole, 
E., 1990, ApJ, 360, 16.

\bibitem[Smail et al.\ 1996a]{s96a}
Smail, I., Dressler, A., Couch, W.J., Ellis, R.S., Oemler, A. Jr., 
Butcher, H.\ \& Sharples, R.M. 1996a, ApJ, in press. (S97a)

\bibitem[Smail et al.\ 1997b]{s97b}
Smail, I., Ellis, R.S., Dressler, A., Couch, W.J., Oemler, A. Jr.,
Sharples, R.M, \& Butcher, H. 1996b, ApJ 479, 70.

\bibitem[Spitzer \& Baade\ 1951]{sb51}
Spitzer, L., and Baade, W. 1951, ApJ, 113, 413.

\bibitem[Whitmore \& Gilmore]{wg91} Whitmore, B.C., \& Gilmore, D.M., 
1991, ApJ, 367, 64.

\bibitem[Whitmore et al.\ 1993]{wgj93} Whitmore, B.C., Gilmore, D.M., \&
Jones, C. 1993, ApJ, 407, 489.

\end{thebibliography}
\end{document}